\documentclass[aps,prd,preprint,groupedaddress,showpacs,nofootinbib]{revtex4}

\usepackage{amsmath}
\usepackage{amssymb}
\usepackage{mathtools}
\usepackage{graphicx}
\usepackage{hyperref}
\usepackage{dsfont}
\usepackage{bbm}

\usepackage{cancel}

\usepackage{enumerate}

\DeclareGraphicsExtensions{.pdf,.png,.jpg,.eps}

\usepackage[utf8]{inputenc}
\usepackage{amsmath}

\begin{document}
\unitlength = 1mm

\title{The role of conformal symmetry in gravity and the standard model}

\author{Stefano Lucat}
\email{s.lucat@uu.nl}

\author{Tomislav~Prokopec}
\email{t.prokopec@uu.nl}

\affiliation{Institute for Theoretical Physics, Spinoza Institute and EMME$\Phi$, Utrecht University,\\
Postbus 80.195, 3508 TD Utrecht, The Netherlands}

\date{\today}

\begin{abstract}
In this paper we consider conformal symmetry in the context of manifolds with general affine connection.
We extend the conformal transformation law of the metric to a general metric compatible affine connection,
and find that it is a symmetry of both the geodesic equation and the Riemann tensor. We derive the generalised
Jacobi equation and Raychaudhuri equation and show that they are both conformally invariant. Using the geodesic
deviation~(Jacobi) equation we analyse the behaviour of geodesics in different conformal frames.

Since we find that our version of conformal symmetry is exact in classical pure Einstein's gravity, we ask whether one
can extend it to the standard model. We find that it is possible to write conformal invariant lagrangians in
any dimensions for vector, fermion and scalar fields, but that such lagrangians are only gauge invariant in four dimensions.
Provided one introduces a dilaton field, gravity can be conformally coupled to matter.
\end{abstract}

\pacs{04.62.+v, 12.60.-i, 98.80.Qc}

\maketitle


\section{Introduction}
\label{Introduction}

Conformal transformations~(Weyl transformations) can be defined as the set of all space-time transformations
that change the line element as,
\begin{equation}
\label{Weyl.rescaling.0} 
 \text{d}s^2 \rightarrow e^{2\theta(x)}  \text{d}s^2 \,.
\end{equation}
The local symmetry group defined by this set can be thought as an extension of the Lorentz group, as follows. The Lorentz group is spanned by the generators
$M_{\mu\nu}= -M_{\nu\mu}$,  which satisfy,
\[\begin{split}
 \big [M_{\mu\nu}, M_{\lambda\sigma} \big ] =& i \left (\eta_{\nu\lambda} M_{\mu\sigma} + \eta_{\mu\sigma} M_{\nu\lambda} - \eta_{\nu\lambda} M_{\mu\sigma} - \eta_{\mu\lambda} M_{\nu\sigma} \right )\,,\\
\bold{\Lambda}(x) =& \exp\left ( M_{\mu\nu} \omega^{\mu\nu}(x)\right )\,,
\end{split}\]
where $\bold{\Lambda}(x)$ is an element of local Lorentz group, {\it i.e.} it represents the transition matrix that relates different (locally)~inertial observers.
Its extension can be thought of as being spanned by the identity matrix $\mathds{1}$ of the representation space where $M_{\mu\nu}$ lives and its exponentiation leads to the transformation $\bold{\Lambda}= \exp\left ({\mathds{1}}\theta(x)\right )$. The additional parameter $\theta(x)$ belongs to $\mathbb{R}$ and,
as we will argue in this paper, extends the notion of Lorentz transformation to include local rescalings of fields.

The transformation law~(\ref{Weyl.rescaling.0}) represents a local rescaling of lengths such that angles and dimensionless ratios are kept the same.
Any theory left invariant under the transformation law~(\ref{Weyl.rescaling.0}), should therefore not possess any intrinsic
length scale, since that would break the symmetry, and all meaningful observables in the theory should be dimensionless
ratios.
The fundamental constants that can be found in nature relate different quantities to each others. For example, the speed of light
is used to relate time and space, $x = c \, t$, the Planck constant relates space and momentum, $\Delta x \Delta p = \hbar$, and the Boltzmann constant
relates energy with temperature, {\it i.e.} $E = k_B T$. The question we want to answer is, do $c,\,\hbar,\,k_B$ rescale if lengths are locally changed as in~(\ref{Weyl.rescaling.0})?
The answer is no: from Lorentz invariance, we know that time and space should be treated equally, hence they should have the same conformal weight under the rescaling~(\ref{Weyl.rescaling.0}). 
This means $c$ is invariant under conformal transformations. We also know that masses scale inversely to length or time, as we can infer by looking at the mass terms in any lagrangian, which implies that 
$\hbar$, whose dimension is $\rm{[kg \, m^2/s]}$ does not rescale if lengths are locally changed as in~(\ref{Weyl.rescaling.0}). Finally, similar arguments can be applied to the Boltzmann constant, 
by knowing that temperature scales as energy. We can therefore conclude that $c,\,\hbar,\,k_B$ relate quantities with the same scaling behaviour under~(\ref{Weyl.rescaling.0}), which means 
that they are conformally invariant. Alternatively put, we cannot construct any intrinsic~(length, time, energy or temperature) scale using just $c,\,\hbar,\,k_B$: any attempt to do so, will produce a constant that relates quantities with the same scaling behaviour 
under the rescaling~(\ref{Weyl.rescaling.0}), and is thus conformally invariant. 
The coupling constants in the Standard Model include the Yukawa couplings  $y_{ij}$ (which are all dimensionless
in natural units, $\hbar=c=k_B=1$, and in four dimensions),
the  couplings of gauge fields to matter fields (charges) $g_i$ (which are also dimensionless)
and the (self-)couplings of scalar fields (which are also dimensionless).
On the other hand the Higgs mass term, which is in the Standard Model a dimensionful coupling parameter,
violates the Weyl
symmetry~(\ref{Weyl.rescaling.0}), and furthermore 
the Newton's constant, $G_N$ ${\rm [m^2]}$, and the cosmological constant, $\Lambda$ ${\rm [m^{-2}]}$, both violate the
symmetry~(\ref{Weyl.rescaling.0}) in the gravitational sector. The question is then how to introduce a scale to
a conformally invariant theory. 

This can be done through 
(matter) fields since they introduce dimensions. Indeed, in four space-time dimensions the canonical dimension of a scalar field $\phi$
is ${\rm [m^{-1}]}$, vector fields $A_\mu$ ${\rm [m^{-1}]}$ and fermionic fields $\psi$ ${\rm [m^{-3/2}]}$.
If matter fields acquire condensates, they can  
introduce a scale in the problem. For example, a scalar field condensate $\langle \phi\rangle$ ${\rm [m^{-1}]}$
and in this paper we shall make use of the dilaton field condensate to introduce the Planck energy scale,
 $\langle \Phi\rangle = M_{\rm P}$ ${\rm [m^{-1}]}$; the same (or analogous) field condensate can set the mass scale of 
 electroweak symmetry breaking $\xi\langle \Phi\rangle$, where $\xi$ is some dimensionless coupling constant.

The scalar that is responsible for the Planck constant may therefore be also responsible for the Higgs mass,
and therefore for all of particle's masses. This way, there would be no absolute intrinsic scale in nature, but only a local 
value of the field condensate $\langle\Phi^2\rangle$.
If we contemplate this possibility, we might ask what differences should local observers perceive due to the fact that 
their masses are generated by field condensates?
Any observer performing local experiments can only observe the local value of the field, and base his or her 
entire system of units upon it. This means in particular that he or she will measure time in units of the field, $\text{d}s^2 = \text{d}\hat{s}^2/\Phi(x)^2$,
and might experience time dilatation if entering a region where the condensate vacuum expectation value (vev) is closer to zero.
If two observers happen to live in different places with different local values of $\langle\Phi^2\rangle$, 
they might meet and compare the results of their experiments,
which might differ quantitatively, but should agree when it comes to dimensionless ratios.

An example of how this might come to be can be found in cosmology, where most observations are based on redshift of photons. 
Since the observed frequency is related to the frequency at emission by,
\[\nu_{\it obs} = \frac{a_{\it em}}{a_{\it obs}}\nu_{\it em}\,,\]
where $a_{\it em}=a(t_{\it em})$ ($a_{\it obs}=a(t_{\it obs})$) 
denotes the cosmological scale factor at the emitter (observer), 
observing the frequency of standard candles can tell us
whether the universe is expanding or contracting, {\it i.e.} whether $a(t)$ is increasing or decreasing. However, as noticed in~\cite{Domenech:2016yxd, Wetterich:2014zta}, there could
be another interpretation: that the masses of particles are increasing as time goes by. If the mass of the atom which emitted the photon
was smaller, the Bohr radius of the atom which emitted it was also smaller, and therefore the emitted frequency will
be measured as if the universe was expanding~\cite{Domenech:2016yxd}. In this interpretation,
the space-time metric can be static, but the masses of particles change, and therefore one observes
redshift in frequency. This just corresponds to use two different conformal frames:
if one choses $\Phi(x)=\Phi_0$ to be the scale today, and constant during the universe evolution, he or she will conclude that the universe is expanding,
while letting $\bar{\Phi}_0(t)$ be changing during the evolution of the universe, leads to the conclusion
that the universe is static, but masses change during its evolution. Note also that the observed and emitted frequency might differ
according to the interpretation, but the dimensionless ratio $\nu_{\it obs}/\nu_{\it em}$ is the same in all frames.

It can be argued that these two interpretations are not really different~\cite{Hooft:2014daa, Wetterich:2014zta}, but are related by a frame transformation of the
type~(\ref{Weyl.rescaling.0}). In the previous example we argued that different scales might be associated to different moments
of the cosmic evolution. Since time and space are on equal footing in gravitational theories, we should expect variations of the energy scale
also in different space regions. From this, we justify the local nature of~(\ref{Weyl.rescaling.0}): the scale used by local observers to perform experiments
can vary locally, and be different in different regions of space-time. If we follow this interpretation, we are lead to consider the local scale
transformation~(\ref{Weyl.rescaling.0}) as a symmetry of Nature, which is broken today, but realised in the fundamental theory.
\begin{figure}
\includegraphics[width=0.5\textwidth]{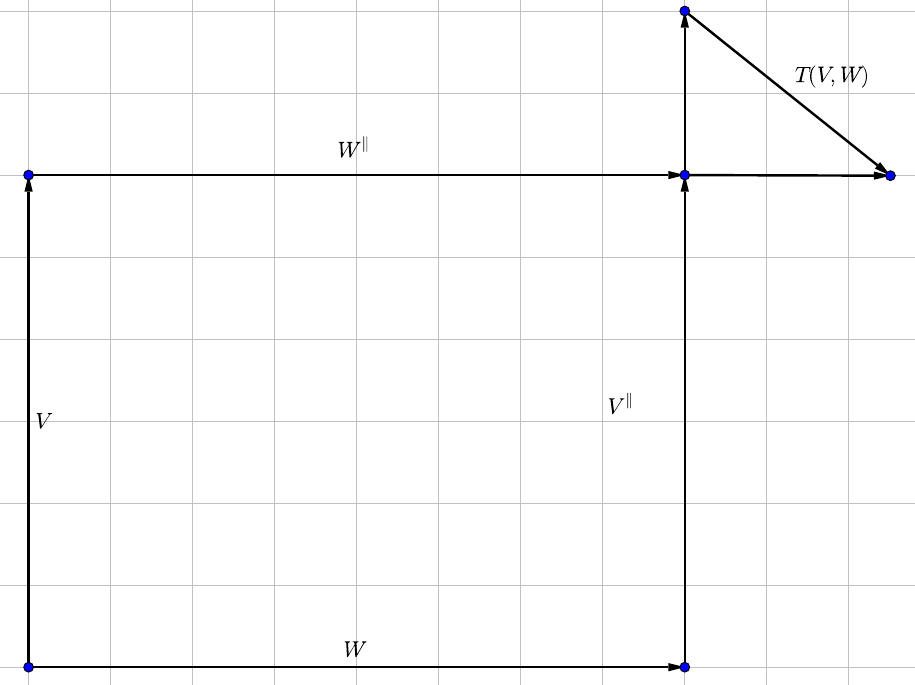}
\includegraphics[width=0.4\textwidth]{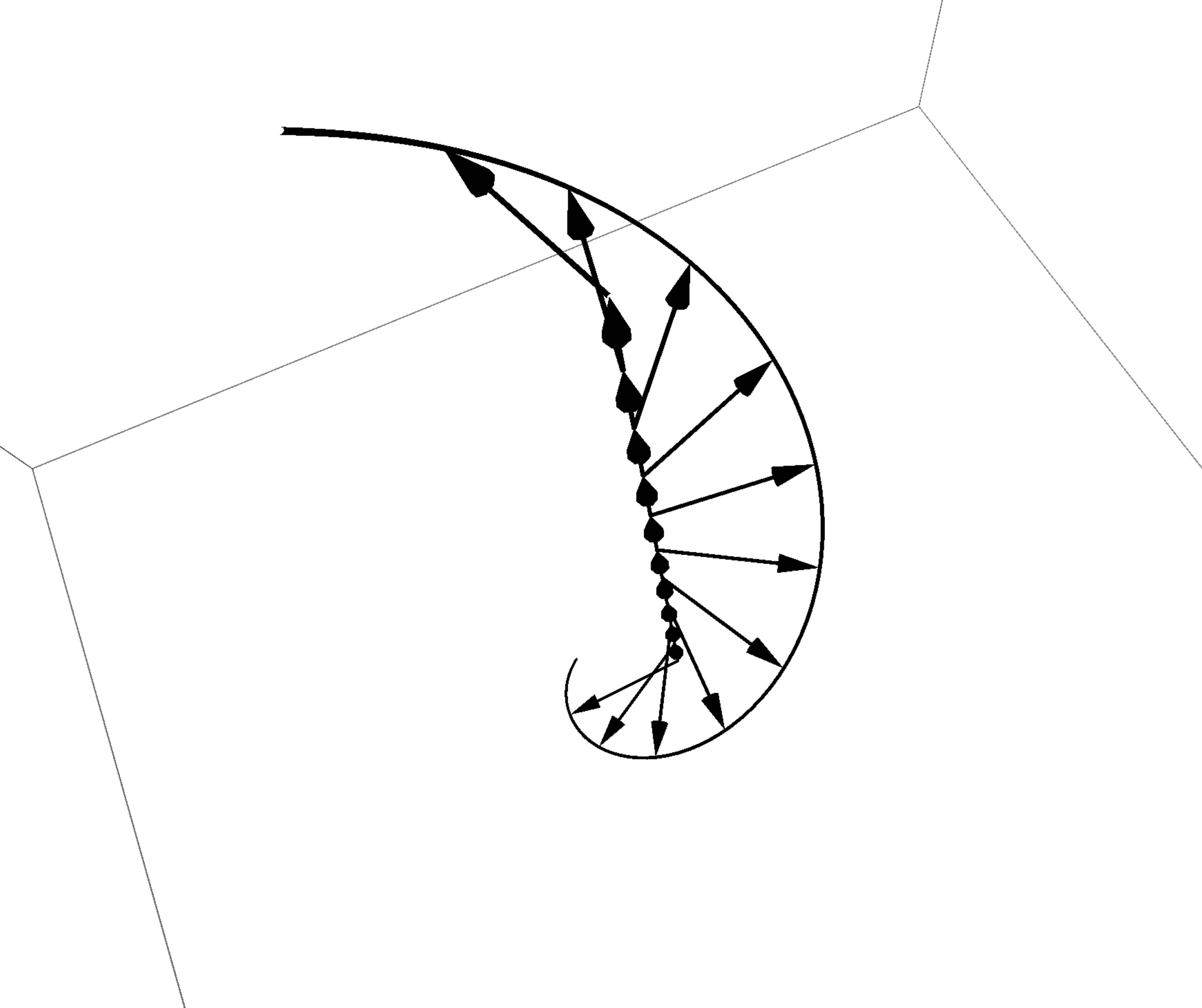}
\caption{On the left, the effect of the torsion trace on parallel transport of vectors: the torsion trace induces a rescaling of vectors during parallel transport.
Note that the parallel transported version of $V$ and $W$ are parallel respectively to $V$ and $W$, and lie on the same plane as their parent vectors.
This feature is characteristic of the torsion trace and does not hold when more components are added to the torsion tensor. On the right, an example of a space
with torsion, and the rotation induced by parallel transport. In the figure we can see a geodesic tangent vector, $\dot{\gamma}$, parallel transported along the vertical direction, and a rotating vector, orthogonal to $\dot{\gamma}$, which is a Jacobi field
in this space. The helicoidal curve traced by the Jacobi field is also a geodesic. }
\label{fig:fig1}
\end{figure}

There are two pieces of evidence that support this conclusion: firstly, the fact that the Standard Model is a nearly conformal theory. Safe for the Higgs mass, all
the interactions of the Standard Model possess dimensionless couplings and safe for the scalar mass term, all other terms of the Standard Model action
are classically conformal. Secondly, it is a well known fact that the renormalization group's equations can possess fixed points~\cite{Zinn-Justin:572813}. These are energy scales at which the coupling constants
of the theory's cease to be scale dependent. If we conjecture that the fundamental theory of nature should possess renormalizability, we have to consider the
existence of an UV fixed point. Near this point, quantum fluctuations will be such to restore conformal symmetry, which would be broken in the IR limit of the theory.
We are therefore lead to consider the transformation~(\ref{Weyl.rescaling.0}) as a symmetry in the UV completion of the fundamental theory, which is then broken
 once the theory flows towards the IR.

The missing piece in this whole argument, however, is gravity. If we want to argue that conformal transformations are on the same footing as coordinate transformations,
{\it i.e.} they merely describe different observers, we should be able to construct a theory of gravity for which the metric rescaling~(\ref{Weyl.rescaling.0}) is a symmetry.
In such a theory, all gravitational observables must be the same in all conformal frames. At the present state of art, there is no theory that is capable of this:
although conformal lagrangians can be constructed using the Weyl tensor or a non minimally coupled scalar~\cite{birrell1984quantum,Shapiro:2001rz}, observables such as curvature and geodesics trajectories
are frame dependent in these theories, and we are therefore forced to strongly modify Einstein's theory of gravity and explain how to flow from the high energy description to
the low energy limit~(Einstein's theory). Furthermore, the gravitational models based on Weyl tensor are generally unstable,
both at the classical and quantum level, 
as they violate Ostrogradsky's theorem~\cite{Woodard:2015zca} and possess ghosts~\cite{Hooft:2010nc}.

In this paper we address these questions in the context of Einstein-Cartan gravity~\cite{RevModPhys.48.393}. In this theory, the only assumption from General Relativity that is dropped is that of a symmetric connection.
The torsion tensor, defined as the antisymmetric part of the connection, is introduced as a novel geometrical notion, distinguished from curvature, and represents
a microscopic twisting of space-time. In a space-time with torsion, vectors rotate during parallel transport, in such a way that parallelograms constructed by parallel transporting
two vectors on each others, do not close~\cite{Nakahara:2003nw}. Vectors also rescale during parallel transport, in a way described by the torsion trace, as is illustrated in figure~\ref{fig:fig1}.
This is the geometrical reason why torsion is linked to conformal symmetry, and its transformation properties under the rescaling~(\ref{Weyl.rescaling.0}) are crucial in the construction
of a frame independent theory of gravity.

In section~\ref{Conformal transformations in general relativity}
we consider the most general metric compatible linear connection, which depends on torsion other than the metric,
and extend the transformation law~(\ref{Weyl.rescaling.0}) to it. We find a remarkably simple
transformation that leaves unaltered the geodesic equation and the Riemann tensor. As a consequence, also the
Einstein's tensor remains invariant under the complete transformation.
Since the other transformations that have the property of leaving geometry invariant are coordinate changes,
and they correspond to a change in the observer's point of view, we argue that the same interpretation should be
put forward for conformal transformations such as~(\ref{Weyl.rescaling.0}).

We then address the question, what different physical properties do different conformal observers
measure? Space-time singularities are defined through geodesic incompleteness: an observer falling into a
singularity would see its proper time stop at a certain point. Clearly, a simple rescaling of the type~(\ref{Weyl.rescaling.0})
can be used to extend the geodesic to arbitrary values of the proper time~\cite{Wetterich:2014zta}. If one wants to follow this interpretation,
he or she can ask what happens to nearby geodesics, as they get closer and closer to the singularity.
We address this question by studying the geodesic deviation~(Jacobi) equation, which we generalise to general space-time
with torsion.
We show that in a different conformal frame, observers will measure a damping force that slows the acceleration
of geodesics towards each others. If the transformation really pushes the singularity to infinite proper time,
this damping force becomes infinite, effectively stopping the force that pulls geodesics towards each others,
when the singularity is reached. The same effect can be described in a less general setting using the Raychaudhuri
equation with torsion, which is also conformally invariant~\footnote{As it should be since a curve's shear, vorticity and divergence, which
appear in the Raychaudhuri equation, are observable quantities, and as such should not depend on the conformal frame used to compute them.}.

In section~\ref{Coupling to matter}, we consider the interaction terms of the standard model,
for fermions, gauge bosons and scalars fields. We show that our extended conformal symmetry can
be easily unified with the Standard Model action, provided that the scalar field kinetic term is modified.
We construct a conformally invariant covariant derivative, using the unique coordinate and
metric independent contraction of torsion as the gauge field to retain local conformal invariance, that is the torsion trace.
In this construction, we treat the torsion trace as a
new ``gauge boson" of the group defined by the transformation~(\ref{Weyl.rescaling.0}), motivated by the fact that the torsion
trace generates local scale transformations, as can be seen in figure~\ref{fig:fig1}.
We show that all interactions in the Standard Model are compatible with
our version of conformal symmetry.
The only way to make the gauge fields action conformally symmetric in $D\neq 4$ is to break
gauge symmetry. If the gauge symmetry is Abelian and in $D=4$ the two local symmetries can be unified at the classical level.

\section{Conformal transformations in general relativity}
\label{Conformal transformations in general relativity}

In cosmology, the transformation~(\ref{Weyl.rescaling.0}) is often used to simplify inflationary models.
For example, a complicated Lagrangian where the inflaton field is non minimally coupled to gravity~(often referred to as Jordan frame), can be studied in a more familiar setting,
by performing the transformation~(\ref{Weyl.rescaling.0}), leading to a minimally coupled theory~(in the Einstein's frame). If the transformation parameter $\theta$ in~(\ref{Weyl.rescaling.0}) is
regular everywhere, classical solutions~\footnote{Since our discussion is fully classical, we do not look in this paper at the quantum behaviour of theories. However, it should be noted and kept in mind that quantum theories break conformal invariance. One reason is that the path integral measure is not invariant under conformal rescaling~(\ref{Weyl.rescaling.0}). Consequently, the well known conformal anomaly~\cite{Duff:1993wm, birrell1984quantum} generates terms which break conformal symmetry.} 
in Jordan frame are mapped onto classical solution in Einstein frame.

However, we can ask whether the two frames are physically equivalent, meaning that any physically meaningful observable
should be the same in both frames. It becomes clear that this is not the case, if we consider the geodesic equation. Since
geodesics are trajectories of free falling bodies, they should be invariant under~(\ref{Weyl.rescaling.0}), if one wants to claim
that the two frames are physically equivalent.

If the metric is locally changed as,
\begin{equation}
\label{Weyl.rescaling} g_{\mu\nu} \rightarrow \tilde{g}_{\mu\nu}= e^{2\theta(x)} g_{\mu\nu} \,, \, \text{d}\tau\rightarrow\text{d}\tilde{\tau}= e^{\theta(x)}\text{d}\tau\,,
\end{equation}
then the Christoffel symbols, $\overset{\circ}{\Gamma}{}^\lambda_{\mu\nu}$, are shifted by,
\begin{equation}
\label{Christoffel.Conformal}
\delta\overset{\circ}{\Gamma}{}^\lambda_{\mu\nu} = \delta^\lambda_{\mu} \partial_{\nu}\theta + \delta^\lambda_{\nu} \partial_{\mu}\theta- g_{\mu\nu}\partial^\lambda\theta\,.
\end{equation}
It then straightforwardly follows that the geodesic equation,
\begin{equation}
\label{zeroth.geodesic.equation}
\frac{\text{d}^2 x^\lambda}{\text{d}\tau^2} + \overset{\circ}{\Gamma}{}^\lambda_{\mu\nu} \dot{x}{}^\mu\dot{x}{}^\nu = 0 \,,
\end{equation}
transforms as,
\begin{equation}
\label{wrong.geodesic.equation} e^{-\theta}\frac{\text{d}}{\text{d}\tilde{\tau}}\left (e^{-\theta}\frac{\text{d}x^\lambda}{\text{d}\tilde{\tau}}\right ) + e^{-2\theta}\left (\overset{\circ}{\Gamma}{}^\lambda{}_{\mu\nu} \dot{x}{}^\mu  \dot{x}{}^\nu +2 \frac{\text{d} \theta}{\text{d}\tilde{\tau}} \dot{x}{}^\lambda - \dot{x}{}^\mu  \dot{x}{}_\mu \partial^\lambda\theta\right )=0\,.
\end{equation}
The third term in Eq.~(\ref{wrong.geodesic.equation}) can be absorbed in a reparametrization of the proper time $\tau$, but
the fourth cannot. This shows that, in absence of torsion, geodesics calculated in two different conformal frames are,
in general, not the same. Clearly this is enough to show that the two conformal frames are not
physically equivalent: since particles cannot follow two orbits at the same time, one has to define which
one is the physical frame, and compute observable quantities in such a frame.

It is of course possible to study a theory by doing the conformal rescaling~(\ref{Weyl.rescaling.0}), but one
has to construct frame independent observables that allow to relate the two frames to each others.
For example, one can construct such observables in theory of cosmological 
perturbations~\cite{Prokopec:2013zya,Prokopec:2012ug,Weenink:2010rr}, such that the observed scalar and
tensor spectra do not depend on whether one calculates them in Einstein or Jordan frame.
In classical General Relativity a conformal observable is the Weyl tensor, the trace-free part of
the Riemann tensor. Such a quantity is indeed invariant under the rescaling~(\ref{Weyl.rescaling.0}),
however, it contains less information than the Riemann tensor itself. Namely, since the Weyl tensor is trace free,
the gravitational scalars used in the Lagrangian of the theory has to be the square Weyl tensor. 
This leads to a theory that is substantially different
from general relativity and would require a sophisticated mechanism to explain why the low energy effective theory of gravity is
Einstein's theory. As we shall see, by adding torsion to the space-time manifold, we can construct conformally invariant theories
using both the Ricci tensor or scalar and the Ricci tensor. 
In some versions of the theory, the modified Einstein's equations are the same one studies in general
relativity, safe for a field-dependent Planck mass
 and as such in the low energy limit they reproduce the same results as general relativity.

We now proceed to derive the transformation laws following the conformal rescaling~(\ref{Weyl.rescaling.0})
in Einstein-Cartan gravity, by demanding that the geodesic equation should be invariant under conformal rescaling.

Let us define the torsion tensor as the antisymmetric part of the connection,
\begin{equation}
\begin{split}
T[X,Y] =&-\frac{1}{2}( \nabla_X Y - \nabla_Y X - [X,Y] )\,, \\
\label{Torsion.Definition.1} \text{or in components, }\,\,T^\lambda{}_{\mu\nu} =&  \Gamma^\lambda{}_{[\mu\nu]} =\frac{1}{2} \left ( \Gamma^\lambda{}_{\mu\nu} -  \Gamma^\lambda{}_{\nu\mu}\right ) \,.
\end{split}
\end{equation}
The Riemann tensor is then,
\begin{equation}
\begin{split}
\label{Riemann.Definition.1}
R[X,Y]Z =& \nabla_X \nabla_Y Z - \nabla_Y \nabla_X Z - \nabla_{[X,Y]} Z \,,\\
 \text{or in components, }\,\,R^\lambda{}_{\sigma\mu\nu}  =& \left ( \partial_\mu \Gamma^\lambda{}_{\sigma\nu}-\partial_\nu \Gamma^\lambda{}_{\sigma\mu} + \Gamma^\lambda{}_{\kappa\mu}\Gamma^\kappa{}_{\sigma\nu} - \Gamma^\lambda{}_{\kappa\nu}\Gamma^\kappa{}_{\sigma\mu} \right )   \,.
\end{split}
\end{equation}
We will denote vectors, forms and tensors both in their components free notation, and as their components in a local basis. Vectors, $V$, act on functions, $f$,
forms, $\omega$, act on vectors, and their action is defined as,
\begin{equation}
\begin{split}
&V[f] = V^\mu\partial_\mu f \,, \text{where $f$ is a function} \,,\\
&\omega[V] = \omega_\mu V^\mu\,, \text{where $V$ is a vector} \,.
\end{split}
\end{equation}
More generally, a tensor $M$ of rank $\binom{p}{q}$, acts linearly on $p$ forms and $q$ vectors to give a real number, as
\begin{equation}
M[\omega_1,\cdots, \omega_p, V_1,\cdots,V_q] = M^{\mu_1\cdots\mu_p}_{\nu_1\cdots\nu_q} \omega_{\mu_1}\cdots\omega_{\mu_p} V^{\nu_1} \cdots V^{\nu_q} \,.
\end{equation}
Finally, for the metric convention, we use the signature $(-,+,+,+)$.

A well known result~\cite{RevModPhys.48.393} is that the most general antisymmetric connection satisfying metric compatibility
is given by,
\begin{equation}
\label{Metric.Compatible.Connection} \Gamma^\lambda{}_{\mu\nu} =T^\lambda{}_{\mu\nu} +  T_{\mu\nu}{}^\lambda + T_{\nu\mu}{}^\lambda + \overset{\circ}{\Gamma}{}^\lambda{}_{\mu\nu}  \,,
\end{equation}
where $\overset{\circ}{\Gamma}{}^\lambda{}_{\mu\nu}$ are the Christoffel symbols computed using the metric, {\it i.e.}
\begin{equation}
\overset{\circ}{\Gamma}{}^\lambda{}_{\mu\nu} = \frac{g^{\lambda\sigma}}{2} \left ( \partial_\mu g_{\sigma\nu} +\partial_\nu g_{\sigma\mu}-\partial_\sigma g_{\mu\nu}\right )\,.
\end{equation}

Let us now consider the geodesic equation: under the conformal rescaling~(\ref{Weyl.rescaling.0}) we have,
\begin{equation}
\begin{split}
\label{geodesic.invariance}&\frac{\text{d} x^\lambda}{\text{d}\tau}\nabla_\lambda\frac{\text{d} x^\mu}{\text{d}\tau} = 0 \rightarrow \frac{\text{d} x^\lambda}{\text{d}\tilde{\tau}}\left (\tilde{\nabla}_\lambda\frac{\text{d} x^\mu}{\text{d}\tilde{\tau}} \right )=\frac{\text{d}^2 x^\mu}{\text{d}\tilde{\tau}^2} + \tilde{\Gamma}{}^\mu{}_{\alpha\beta}\frac{\text{d} x^\alpha}{\text{d}\tilde{\tau}}\frac{\text{d} x^\beta}{\text{d}\tilde{\tau}} = \\
=& e^{-2\theta} \left ( \frac{\text{d}^2x^\mu}{\text{d}\tau^2} +( \Gamma^\mu{}_{\alpha\beta} +\delta\Gamma^\mu{}_{\alpha\beta}) \dot{x}{}^\alpha\dot{x}{}^\beta - \dot{\theta} \dot{x}{}^\mu \right ) = 0\,,
\end{split}
\end{equation}
where, inspired by~(\ref{Christoffel.Conformal}), we postulated that the connection transforms linearly,
\[{\Gamma}{}^\mu{}_{\alpha\beta}\rightarrow\tilde{\Gamma}{}^\mu{}_{\alpha\beta} = {\Gamma}{}^\mu{}_{\alpha\beta}+\delta{\Gamma}{}^\mu{}_{\alpha\beta}\,.
\]
We see from Eqs.~(\ref{Christoffel.Conformal}), 
(\ref{Metric.Compatible.Connection}) and~(\ref{geodesic.invariance}) that the most natural choice is to write,
\begin{equation}
\label{transformation.law.connection}  \delta\Gamma^\mu_{\alpha\beta} = \delta^\mu_\alpha \partial_\beta \theta \,, \qquad\delta T^\mu{}_{\alpha\beta} =\delta^\mu_{[\alpha} \partial_{\beta]} \theta\,, \qquad2\delta T_{(\alpha\beta)}{}^\mu = g_{\alpha\beta}\partial^\mu \theta - \delta_{(\alpha}^\mu \partial_{\beta)} \theta \,,
\end{equation}
such that the geodesic equation is mapped onto itself. The transformation law~(\ref{transformation.law.connection}) has been considered in the literature before, for example in~\cite{Maluf:1985fj,Shapiro:2001rz},
where the authors consider coupling scalar-tensor theories to torsion. In~\cite{Fonseca-Neto2013}, the authors find the existence of an equivalent class of manifolds, analogous to a metric $e^{2\theta}\hat{g}_{\mu\nu}$,
and torsion purely given by $\delta T^\mu{}_{\alpha\beta}$, as in~(\ref{transformation.law.connection}), and claim that they are different representation of general relativity.
We pursue such interpretation, in this paper, as a manifestation of invariance of physical observables for different observers, which is reflected in the fact that the geometry
remains invariant under the symmetry. For the simpler case studied in~\cite{Fonseca-Neto2013}, the theory is indeed analogous to General Relativity,
but in the general case, one has to consider the torsion trace as an external field. We show in section~\ref{Coupling to matter} that, in the classical limit,
the case of pure gauge torsion~(\ref{transformation.law.connection}) is a solution of the theory, and interpret the scalar parameter in~\cite{Fonseca-Neto2013}
as the dilaton which sets the Planck scale.

The conformal transformations~(\ref{transformation.law.connection}) map geodesic trajectories onto geodesic trajectories in the new frame,
acting as a reparametrisation of the proper lengths. This is the case if torsion is included in the geodesic equation.
In principle, one can choose not to transform the torsion tensor, and just transform the Christoffel connection
as in Eq.~(\ref{Christoffel.Conformal}).
However, in our opinion,~(\ref{transformation.law.connection}) is the most natural choice: namely,
from differential geometry we know that a tensor is constant along the integral curves of a vector field $X$ if,
\begin{equation}
\label{parallel.transport.1}\nabla_X T = 0 \,,
\end{equation}
which, if the tensor acts on $p$ 1-forms and on $q$ vectors, transforms under conformal rescaling as,
\begin{equation}
\label{parallel.transport.conformal}(\nabla_X T + (p-q) X[\theta] T) = 0 \,.
\end{equation}
Requiring invariance of the parallel transport equation~(\ref{parallel.transport.1}) leads to the following transformation's law for
tensors of rank $\binom{p}{0}$ and $\binom{0}{p}$ and their covariant derivatives,
\begin{equation}
\begin{split}
\label{tensors.covariant.transformation.law}\binom{p}{0}: \,\tilde{T}{}^{\alpha_1\cdots\alpha_p} = e^{-p\theta}T^{\alpha_1\cdots\alpha_p}\,, \qquad\nabla_\mu T^{\alpha_1\cdots\alpha_p} \rightarrow& \tilde{\nabla}_\mu \tilde{T}{}^{\alpha_1\cdots\alpha_p} = e^{-p\theta} \nabla_\mu T^{\alpha_1\cdots\alpha_p} \,\,; \\
\binom{0}{p}: \, \tilde{W}{}_{\alpha_1\cdots\alpha_p} = e^{p\theta}W_{\alpha_1\cdots\alpha_p}\,,\qquad\nabla_\mu W_{\alpha_1\cdots\alpha_p} \rightarrow& \tilde{\nabla}_\mu \tilde{W}{}_{\alpha_1\cdots\alpha_p} = e^{p\theta} \nabla_\mu W_{\alpha_1\cdots\alpha_p}\,\,.
\end{split}
\end{equation}
Following these rules, we have that all scalar contractions of tensors are invariant
under conformal transformations. This choice is the most natural from the geometrical perspective,
however it does not give the correct prescription when looking at fields. As an example, consider a
scalar field, which under conformal transformations~(\ref{Weyl.rescaling.0} ) changes as,
\begin{equation}
\label{Scalar.Field.Conformal.Transformation} 
 \phi(x) \rightarrow \tilde{\phi}(x) = e^{-\frac{D-2}{2} \theta} \phi(x) \,,
\end{equation}
while the geometrical prescription would give $\phi\rightarrow \phi$, since $\phi$ is a geometric scalar function.
This does not, however, constitute a problem: scalar fields are in general different objects than geometrical
scalar functions, and can therefore posses different scaling properties, even if their transformation law
under coordinate transformation are the same. 
Scalar fields such as $\phi$ in~(\ref{Scalar.Field.Conformal.Transformation}) can be considered as densitized 
geometric scalars, $\phi=|g|^{x}\phi_g$, where $\phi_g$ denotes a geometric scalar (which does
not transform) and $x=x(w_\phi)=w_\phi/(2D)$ is a function of the conformal weight $w_\phi$ 
of the field $\phi$ (in the above example, $w_\phi=-(D-2)/2$).

We will see in the next section how to construct a conformal invariant covariant derivative
for a field of general conformal weight $w$. For the time being, we focus on
tensors and forms that transform as direct product of 4-velocities, for which the property~(\ref{tensors.covariant.transformation.law})
holds.

The second main property of the connection transformation law~(\ref{transformation.law.connection}) is that
it leaves the Riemann tensor unchanged, as we can see from,
\begin{equation}
\begin{split}
\label{Riemann.Invariance}\tilde{R}{}^\lambda{}_{\sigma\mu\nu} =&\big (\partial_\mu \Gamma^\lambda_{\sigma\nu} + \delta^\lambda_\sigma \partial_\mu\partial_\nu \theta +  \delta^\lambda_\sigma \partial_\mu\theta\partial_\nu \theta -  \partial_\nu \Gamma^\lambda_{\sigma\mu}-\delta^\lambda_\sigma \partial_\nu\partial_\mu \theta- \delta^\lambda_\sigma \partial_\mu\theta\partial_\nu \theta  \\
&+\Gamma^\lambda_{\kappa\mu}\Gamma^\kappa_{\sigma\nu} + \Gamma^\lambda_{\sigma\mu}\partial_\nu\theta + \Gamma^\lambda_{\sigma\nu}\partial_\mu\theta + \delta^\lambda_\sigma \partial_\mu\theta\partial_\nu\theta \\
&- \Gamma^\lambda_{\kappa\nu}\Gamma^\kappa_{\sigma\mu} -\Gamma^\lambda_{\sigma\nu}\partial_\mu\theta - \Gamma^\lambda_{\sigma\mu}\partial_\nu\theta - \delta^\lambda_\sigma \partial_\nu\theta\partial_\mu\theta \big )=  R^\lambda{}_{\sigma\mu\nu}\,.
\end{split}
\end{equation}
An immediate consequence of Eq.~(\ref{Riemann.Invariance}) is that the geometrical side of Einstein's equations
is invariant under conformal transformation the way we have defined them here, that is,
\[
G_{\mu\nu}=R_{\mu\nu} - \frac{1}{2} g_{\mu\nu} R \rightarrow\tilde{R}{}_{\mu\nu} - \frac{1}{2} \tilde g_{\mu\nu} \tilde{R} = R_{\mu\nu} - \frac{1}{2} g_{\mu\nu} R\,.\]
However, the right hand side of Einstein's equation, the matter side, does not have the same property.
In fact, the scaling properties of the energy momentum tensor, in $D$ dimensions, follows from its definition, if $S_{\it m}$ is conformally invariant,
\begin{equation}
\label{Scaling.Dimension.Energy.Tensor} T_{\mu\nu} = -\frac{2}{\sqrt{-g}} \frac{\delta S_{\it m} }{\delta g^{\mu\nu}} \rightarrow \tilde{T}_{\mu\nu} = e^{-(D-2)\theta} T_{\mu\nu}\,,
\end{equation}
since $\sqrt{-g} \delta g^{\mu\nu} \rightarrow e^{(D-2)\theta}\sqrt{-g} \delta g^{\mu\nu}$.
The most simple way of making Einstein's equations conformally invariant, is to write a scalar field~(dilaton) in Einstein's equation, playing the role of a coupling
constant,
\begin{equation}
\label{Conformal.Einstein.Equations} 
R_{\mu\nu} - \frac{1}{2}g_{\mu\nu} R = \frac{\alpha^2}{\Phi^2(x)} T_{\mu\nu}\,,
\end{equation}
where $\alpha^2>0$ is a dimensionless coupling constant and conformal weight of $\Phi$ 
is $w_\Phi=-(D-2)/2$. 
Equations~(\ref{Conformal.Einstein.Equations}) are conformal in any dimension, 
and they follow from the conformal invariant actions, $S_{CG}=\int \text{d}^D x \sqrt{-g}\Phi^2 R$ and $S_m$.
This of course implies that the Newton constant is field dependent, and its apparent value today needs to be generated
by a dilaton condensate, perhaps in a similar way as the Higgs mechanism generates masses in the standard model.
We will return on this issue in section~\ref{Conclusion}, where we discuss possible mechanisms that can lead to spontaneous breaking
of conformal symmetry. For the moment, let us analyse in more depth the consequences of~(\ref{geodesic.invariance}--\ref{Riemann.Invariance}).

In general relativity, the equation that describes the acceleration or deviation of nearby geodesics
is the Jacobi equation,
\begin{equation}
\label{Standard.Jacobi.equation} 
\nabla_{\dot{\gamma}}\nabla_{\dot{\gamma}} J = R[\dot{\gamma},J] \dot{\gamma} \,,
\end{equation}
where $J$ are Jacobi vector fields, $\dot{\gamma}$ is the tangent vector to the geodesic and $R[\dot{\gamma},J] \dot{\gamma}$
denotes the Riemann tensor.

We want to derive the equation corresponding to~(\ref{Standard.Jacobi.equation}) in the framework of gravity with torsion
and study its behaviour in different conformal frames. To this end one can select a bunch of points along the integral curves
of the vector field $J$ and look at the geodesics that start from such points. These in general relativity describe the trajectories of freely falling
particles, and the geodesic deviation equation, that is the analogue of~(\ref{Standard.Jacobi.equation}) will describe the acceleration of such
test bodies towards each others due to the action of gravity.

To look at the way geodesics are pulled towards each other, we construct a variation of geodesics, that is, a set of curves
$\Gamma(\tau,\sigma)$, such that for fixed $\sigma$, $\Gamma(\tau, \sigma_0)$ is a geodesic.
Then we define the Jacobi field and the geodesic tangent vector as,
\begin{equation}
\begin{split}
\label{Jacobi.Field.Definition} 
J =& \frac{\partial \Gamma(\tau,\sigma)}{\partial \sigma }\bigg |_{\tau= 0}\,,
\dot{\gamma} = \frac{\partial \Gamma(\tau,\sigma)}{\partial \tau }\bigg |_{\sigma = 0}\,.
\end{split}
\end{equation}
It follows from these definitions that $J$ and $\dot{\gamma}$ form coordinate lines, and therefore~\cite{Hawking:1973uf},
\[\mathcal{L}_{\dot{\gamma}} J = [\dot{\gamma},J]=  0 \,,\]
where $\mathcal{L}$ is the Lie derivative, and $[\cdot,\cdot]$ denotes the commutator.
We then have that the covariant derivatives of $J$ and $\dot{\gamma}$ satisfy,
\begin{equation}
\label{Parallel.transport} \nabla_J \dot{\gamma} = \nabla_{\dot{\gamma}} J - 2 T[J, \dot{\gamma}] \,,
\end{equation}
as follows straightforwardly from the torsion definition~(\ref{Torsion.Definition.1}). Next, by taking the covariant derivative with
respect to $\dot{\gamma}$ of Eq.~(\ref{Parallel.transport}), and applying the definition of the Riemann tensor~(\ref{Riemann.Definition.1}),
one finds,
\[\begin{split}
&\nabla_{\dot{\gamma}} \nabla_{\dot{\gamma}} J - 2 \nabla_{\dot{\gamma}}T[J, \dot{\gamma}] = \nabla_{\dot{\gamma}} \nabla_J \dot{\gamma} = \\
=&\nabla_J  \nabla_{\dot{\gamma}} \dot{\gamma} + \big [\nabla_{\dot{\gamma}} , \nabla_J \big ] \dot{\gamma} =  \big [\nabla_{\dot{\gamma}} , \nabla_J \big ] \dot{\gamma} - \nabla_{[\dot{\gamma},J]} \dot{\gamma} = R[\dot{\gamma},J]\dot{\gamma}\,.
\end{split}
\]
We thus find that,
\begin{equation}
\label{Jacobi.Equation}  \nabla_{\dot{\gamma}} \nabla_{\dot{\gamma}} J + 2 \nabla_{\dot{\gamma}} T[\dot{\gamma},J] = R[\dot{\gamma},J]\dot{\gamma} \,,
\end{equation}
which is the Jacobi equation for space-times with torsion, and it is the correct generalisation of Eq.~(\ref{Standard.Jacobi.equation}).

We will now demonstrate that equation~(\ref{Jacobi.Equation}) is conformally invariant.
\begin{figure}
\includegraphics[width=0.48\textwidth]{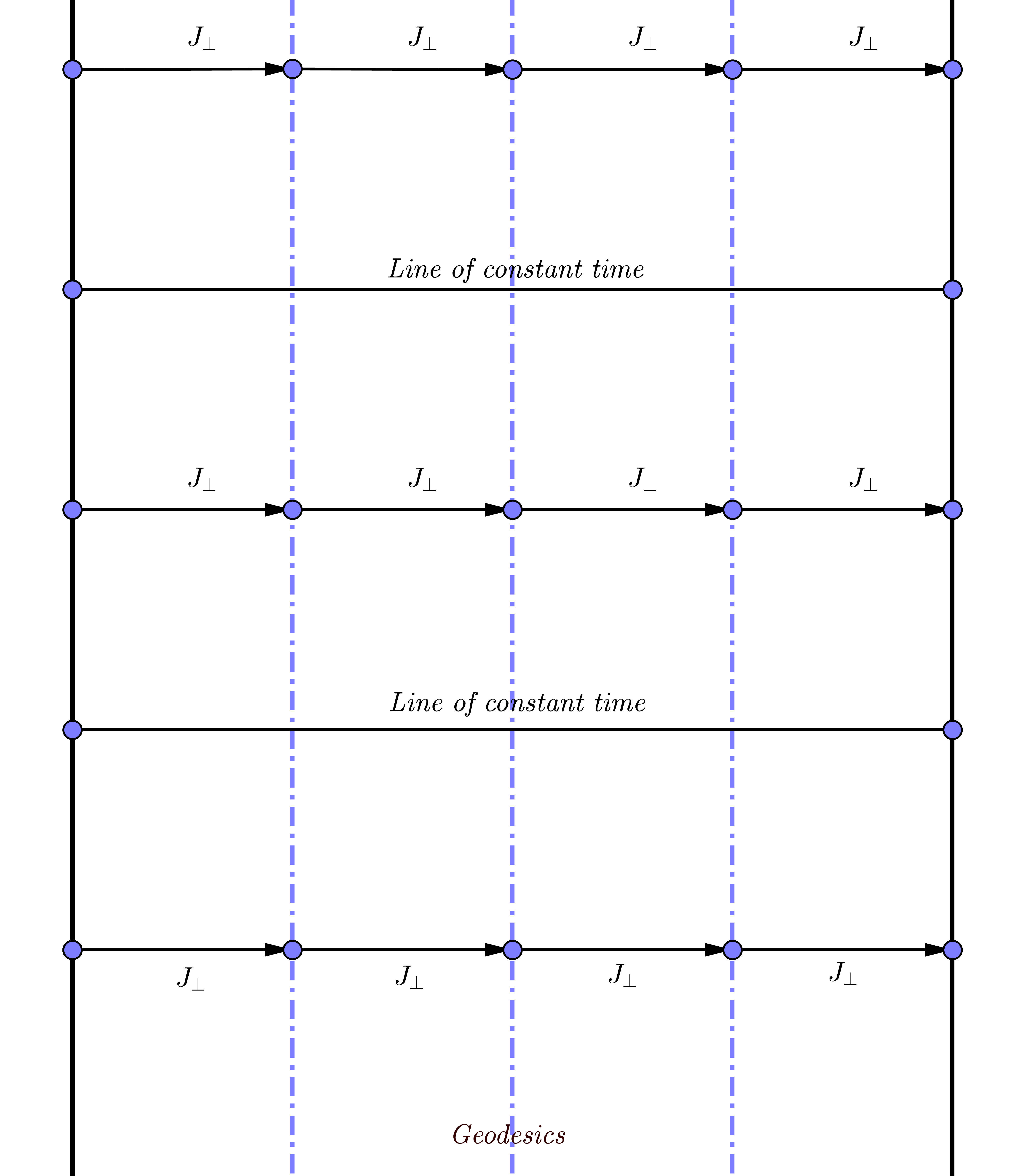}
\includegraphics[width=0.48\textwidth]{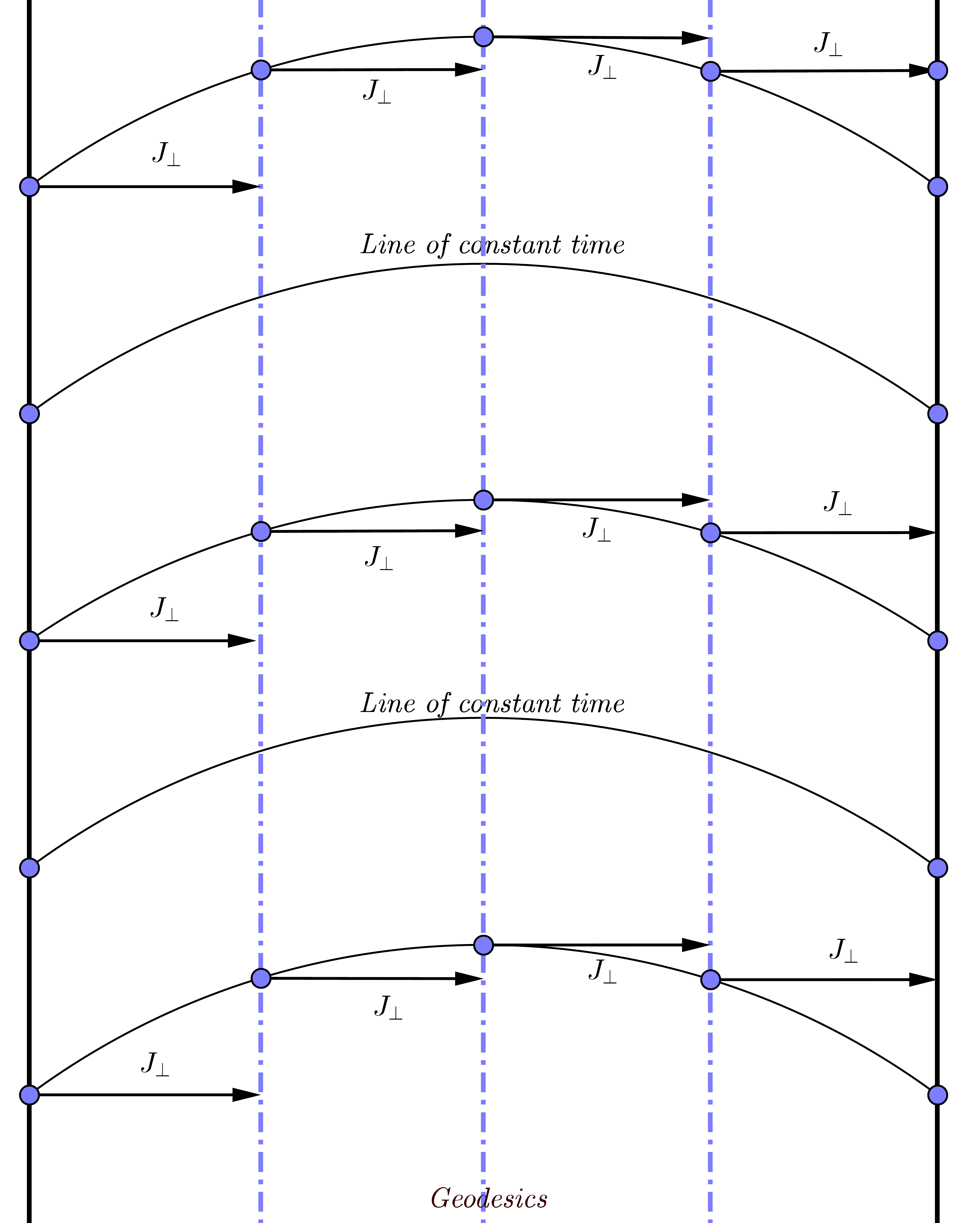}
\caption{Geodesics in two different conformal frames for the flat plane: in both frames geodesics are straight lines, but on the right the equal time lines are bended. However, the component of $J_\perp$ remain unchanged, and are kept constant on the geodesics. }
\label{fig:fig2}
\end{figure}
However, before we proceed, we remind that there exist two kinds of Jacobi fields:
in the direction of $\dot{\gamma}$, there are always two linearly independent solutions, $\dot{\gamma}$ and $\tau \dot{\gamma}$,
as one can easily verify by applying to them Eq.~(\ref{Jacobi.Equation}).
The second fact to be noticed, is that one can project Eq.~(\ref{Jacobi.Equation}) onto the subspace orthogonal to $\dot{\gamma}$,
since the projector operator, $h^\mu_\nu = \delta^\mu_\nu - \epsilon \dot{\gamma}^\mu\dot{\gamma}_\nu$~\footnote{Here $\epsilon=g(\dot{\gamma},\dot{\gamma})$. Note that
$h^\mu_\nu$ is only well defined for time-like and space-like geodesics, since for null geodesics it would give the identity. This happens
because $h_\nu^\mu$ is degenerate for null hypersurfaces. Our construction, and the consequent Raychaudhuri equation, can be straightforwardly generalised to null geodesics, following the
steps in~\cite{Hawking:1973uf}.}, commutes with the differential operator
of Eq.~(\ref{Jacobi.Equation}). By splitting $ J =( \alpha +\beta \tau) \dot{\gamma} + {J}_\perp$, one finds that $T[\dot{\gamma},J] = T[\dot{\gamma},J_\perp]$,
and the same is true for the right-hand side, since both $T[X,Y]$ and $R[X,Y]X$ are antisymmetric under the exchange of $X$ and $Y$.

Therefore, projecting the Jacobi equation~(\ref{Jacobi.Equation}) on the  subspace orthogonal to geodesics leads to,
\begin{equation}
\label{jacobi.equation.orthogonal.projection} \nabla_{\dot{\gamma}} \nabla_{\dot{\gamma}} J_\perp +2 \nabla_{\dot{\gamma}} T_\perp[ \dot{\gamma},J_\perp] = R_\perp[\dot{\gamma},J_\perp]\dot{\gamma} \,,
\end{equation}
where $T_\perp( \dot{\gamma},J_\perp) =( h^\mu_\nu T^\nu{}_{\alpha\beta} \dot{\gamma}^\alpha J^\beta )\partial_\mu$, is the
projection of torsion on the hyperspace perpendicular to the geodesic.
Note that, since the Riemann tensor is antisymmetric in its first two indices, we have $g(\dot{\gamma}, R[\dot{\gamma},J_\perp]\dot{\gamma} ) =0$,
which implies that projecting the right hand side of~(\ref{Jacobi.Equation}) is irrelevant, since $R[\dot{\gamma},J_\perp]\dot{\gamma} = R_\perp[\dot{\gamma},J_\perp]\dot{\gamma}$.

Conformal transformations are essentially reparametrisations of the proper time,
$\text{d}\tau\rightarrow e^{\theta}\text{d}\tau$, $\text{d}\tau^2=-\text{d}s^2$. The integral lines of $J$ represents lines
of constant time on the neighbouring geodesics. It follows from this, that reparametrisations of proper time only change the component
of $J$ in the direction of the geodesics itself, while $J_\perp$ should stay invariant.
We then postulate the following transformation laws for the Jacobi field,
\begin{eqnarray}
\label{Jacobi.Field.Transformation.Law} J_\perp &\rightarrow& J_\perp \,, \\
\dot{\gamma}&\rightarrow& e^{-\theta}\dot{\gamma} \,,\\
\tau\dot{\gamma}&\rightarrow& \tilde{\tau}e^{-\theta}\dot{\gamma}\,, \tilde{\tau}= \int_{\tilde{\tau}_0}^{\tilde{\tau}} e^{-\theta(x(s))}\text{d}s\,.
\end{eqnarray}
This is a consistent choice, because $J_\perp$ and $\dot{\gamma}$ are different geometrical objects: the first contains information about the separation
between different freely falling observers, while the second is the four-velocity of these point-like observers. Therefore, we should not be surprised that the two
vectors possess different scaling properties. However, we should notice that the magnitude of $J_\perp$ is not invariant, contrary to what happens to $\dot{\gamma}$.
In fact, $g(J_\perp,J_\perp)\rightarrow e^{2\theta}g(J_\perp,J_\perp)$, which implies that the measured magnitude
 in one frame, $\| J_\perp\|$,
can be arbitrarily smaller than the measured magnitude in another frame.

We can now show that the Jacobi equation~(\ref{jacobi.equation.orthogonal.projection}) is conformally invariant. In a different frame we would write, for the left hand side of equation~(\ref{jacobi.equation.orthogonal.projection}),
\begin{eqnarray}
\label{left.hand.side.Jacobi} &&\tilde{\nabla}_{\tilde{\dot{\gamma}}}  \tilde{\nabla}_{\tilde{\dot{\gamma}}} J_\perp +2  \tilde{\nabla}_{\tilde{\dot{\gamma}}} \tilde{T}_\perp[ \tilde{\dot{\gamma}},J_\perp]  =\\
 &=&e^{-\theta}\nabla_{\dot{\gamma}}\left [e^{-\theta}\left (  \nabla_{\dot{\gamma}} J_\perp+\dot{\theta}J_\perp \right ) \right ] +e^{-\theta}\dot{\theta}\left [e^{-\theta}\left (  \nabla_{\dot{\gamma}} J_\perp+\dot{\theta}J_\perp \right ) \right ] \nonumber\\
 &&+2 e^{-\theta}\nabla_{\dot{\gamma}} \left[e^{-\theta}\left (T_\perp[ {\dot{\gamma}},J_\perp] -\frac{1}{2}\dot{\theta}J_\perp\right )\right ]+2  e^{-\theta}\dot{\theta}\left (T_\perp[ \tilde{\dot{\gamma}},J_\perp] -\frac{1}{2}\dot{\theta}J_\perp\right )  \nonumber\\
  &=&e^{-2\theta}\left (\nabla_{\dot{\gamma}} \nabla_{\dot{\gamma}} J_\perp +2 \nabla_{\dot{\gamma}} T_\perp[ \dot{\gamma},J_\perp]\right ) 
  \,,
 \end{eqnarray}
and for the right hand side we have,
 \begin{eqnarray}
\label{right.hand.side.Jacobi} \tilde{R}[\tilde{\dot{\gamma}},\tilde{J}{}_\perp]\tilde{\dot{\gamma}}= e^{-2\theta}R[\dot{\gamma},J_\perp]\dot{\gamma}  \,,
\end{eqnarray}
implying that both side of equation scale the same way, thus rendering the Jacobi equation~(\ref{jacobi.equation.orthogonal.projection})
conformally invariant. This is precisely 
what we expected from the conformal invariance of the geodesic equation and of the Riemann tensor.

From the Jacobi equation~(\ref{jacobi.equation.orthogonal.projection}), we can easily derive the Raychaudhuri equation,
which has a more physically intuitive interpretation, borrowed from the context of fluid dynamics. Defining the shear, vorticity and divergence~(or local expansion rate) of geodesics as,
\begin{eqnarray}
\label{shear} S_{\mu\nu} &=& \frac{1}{2} \left (\nabla_\mu \dot{\gamma}_\nu + \nabla_\nu \dot{\gamma}_\mu\right ) \,,\\
\label{vorticity} A_{\mu\nu} &=& \frac{1}{2} \left (\nabla_\mu \dot{\gamma}_\nu - \nabla_\nu \dot{\gamma}_\mu\right ) \,,\\
\label{divergence} \Theta &=& h^\mu_\nu \nabla_\mu \dot{\gamma}^\nu = \nabla_\nu \dot{\gamma}^\nu
\,,
\end{eqnarray}
we can obtain the Raychaudhuri equation by isolating the $J_\perp$ dependence in~(\ref{jacobi.equation.orthogonal.projection}).
Defining the tensor $\Pi_\mu{}^\nu \equiv \nabla_\mu \dot{\gamma}^\nu$ as the covariant derivative of geodesic tangent vector, we find it satisfies,
\[
\nabla_{\dot{\gamma}} \Pi_\mu{}^\nu = -\left (\Pi_\mu{}^\sigma \Pi_\sigma{}^\nu -2 T^\alpha{}_{\sigma\mu} \dot{\gamma}^\sigma  \nabla_\alpha \dot{\gamma}^\nu + R^\nu{}_{\sigma\mu\lambda} \dot{\gamma}^\sigma \dot{\gamma}^\lambda\right ) \,,
\]
of which we can take the trace, to obtain the equation for $\Theta$, as defined in~(\ref{divergence}), in terms of vorticity and shear~(\ref{shear}--\ref{vorticity}),
\begin{equation}
\label{Raychaudhuri.Equation} \frac{\text{d} \Theta }{\text{d}\tau} 
= \left ( 2 A_{\mu\nu}A^{\mu\nu} - 2 S_{\mu\nu}S^{\mu\nu} - \frac{\Theta^2}{3} - R_{\mu\nu} \dot{\gamma}^\mu \dot{\gamma}^\nu 
+2 T^\alpha{}_{\beta\delta}  \dot{\gamma}^\beta  \nabla_\alpha \dot{\gamma}^\delta \right )\,.
\end{equation}
Note that the Raychaudhuri equation~(\ref{Raychaudhuri.Equation}) is conformal, as we can verify by using the transformation
laws of vorticity, shear and local expansion rate,
\begin{equation}
S_{\mu\nu} \rightarrow e^\theta S_{\mu\nu}  \,,\qquad
 A_{\mu\nu} \rightarrow e^\theta A_{\mu\nu} \,,\qquad
 \Theta \rightarrow e^{-\theta} \Theta \,,
 \end{equation}
applying the transformation law for $T^\alpha{}_{\beta\delta}$,
\begin{equation}
\begin{split}
2 T^\alpha{}_{\beta\delta}  \dot{\gamma}^\beta  \nabla_\alpha \dot{\gamma}^\delta \rightarrow 
&e^{-2\theta} \left (2 T^\alpha{}_{\beta\delta}  \dot{\gamma}^\beta  \nabla_\alpha \dot{\gamma}^\delta 
+ \partial_\delta \theta \dot{\gamma}^\alpha  \nabla_\alpha \dot{\gamma}^\delta 
- \dot{\gamma}^\beta\partial_\beta \theta \delta^\alpha_\delta \nabla_\alpha \dot{\gamma}^\delta\right ) =  \\
\label{Torsion.acting.covariant.derivative.tangent.vector}
=&e^{-2\theta} \left (2 T^\alpha{}_{\beta\delta}  \dot{\gamma}^\beta  \nabla_\alpha \dot{\gamma}^\delta 
- \frac{\text{d}\theta}{\text{d}\tau} \, \Theta\right ) \,,
\end{split}
\end{equation}
and noticing that the last term in Eq.~(\ref{Torsion.acting.covariant.derivative.tangent.vector}) cancels against the term coming from transforming the left hand side of Eq.~(\ref{Raychaudhuri.Equation}), namely,
\begin{equation}
\label{equation not important}
\frac{\text{d} \tilde{\Theta} }{\text{d}\tilde{\tau}} 
=e^{-\theta} \frac{\text{d} \left (e^{-\theta}\Theta \right )}{\text{d}\tau} = e^{-2\theta} \left (\frac{\text{d}\Theta}{\text{d}\tau} - \frac{\text{d}\theta}{\text{d}\tau}\, \Theta\right ) \,.
\end{equation}

Furthermore, the fluid vorticity, shear and divergence are observables: they scale commensurably, with a conformal weight of $-1$.
In case of global cosmological space-times~(Friedmann space-times), we have that $\Theta = (D-1) H$, where $H$ is the Hubble rate.
Any cosmological measurement that intends to measure the (global) expansion rate $H(t)$ can in fact only measure the local expansion 
rate $\Theta(x)$ since measurement
are performed locally, in the vicinity of the Earth. $\Theta$ is an observable only if geodesics and $\Theta$ 
are computed using the covariant derivative with torsion,
which provides further theoretical support in favour of the approach proposed in this paper.

Anticipating section~\ref{Gravity plus dilaton: a toy model}, in which we study conformal gravity  
endowed with a dilaton field $\Phi$ and coupled to conformal matter  where we show that, when  
torsion is in the so-called pure gauge form, 
the metric $\text{d}s^2$ can be written as in Eq.~(\ref{Local.Observers.Scale}), such that 
conformal transformation of $\text{d}s^2$ can be obtained by transforming the dilaton field alone as, 
\begin{equation}
\text{d}\ln(\Phi)=-\text{d}\theta
.
\label{equation very important}
\end{equation}
When this is inserted into the Raychaudhuri equation~(\ref{Raychaudhuri.Equation})
one obtains,
\begin{equation}
\begin{split}
\label{General.Relativity.Equation}
\frac{\text{d} \Theta }{\text{d}\tau} =& 
\left ( 2 A_{\mu\nu}A^{\mu\nu} - 2 S_{\mu\nu}S^{\mu\nu} - \frac{\Theta^2}{3} - R_{\mu\nu} \dot{\gamma}^\mu \dot{\gamma}^\nu-\dot{\theta}\, \Theta \right ) \\=& 
\left (2 A_{\mu\nu}A^{\mu\nu} - 2 S_{\mu\nu}S^{\mu\nu} -\frac{\Theta^2}{3} - R_{\mu\nu} \dot{\gamma}^\mu \dot{\gamma}^\nu +\frac{\dot{\Phi}}{\Phi}\, \Theta \right )\,,
\end{split}
\end{equation}
%
where in the first line we took account of Eq.~(\ref{equation not important}) and the second line is obtained 
from~(\ref{equation very important}).
We stress that the quantities $\Theta\,, A_{\mu\nu}\,,S_{\mu\nu}$, appearing in Eq.~(\ref{General.Relativity.Equation}) 
are exactly mapped in the one computed in general relativity,
when torsion is in the pure gauge form from Eq.~(\ref{transformation.law.connection}). In this case Einstein's general relativity
endowed with a (Brans-Dicke) scalar field 
corresponds to a specific gauge choice of a more general theory with torsion in which the torsion tensor can be made 
to vanish identically by a suitable gauge choice.

Eq.~(\ref{General.Relativity.Equation}) describes the behaviour of neighbouring geodesics in different conformal frames and is used to find conditions for which singularities form~\footnote{Singularities are essentially points in which $\Theta\rightarrow -\infty$.}.
First note that in this setting, if $A_{\mu\nu}=0$ in one conformal frame, it will be zero in every frame, since it changes
as $A_{\mu\nu}\rightarrow e^\theta A_{\mu\nu}$. Therefore vorticity cannot prevent conjugate points to form,
if torsion is in its pure gauge form from Eq.~(\ref{transformation.law.connection}). However, because of conformal invariance, we can always switch to a different frame, where
the term $\propto\dot{\Phi}/\Phi$ can slow down convergence of geodesics, which might prevent the formation of conjugate points.
This would mean that singularities can be moved to infinite proper time, in a different conformal frame, and if we argue that such a frame
is the physical frame used by freely falling observers, we would conclude that space-time singularities cannot be reached by any physical
observer.

In fact, singularities might be just bad choices of the conformal frame used: analogously to coordinate transformation,
conformal transformations can be singular, and well defined only in local patches of the space-time manifold~(as for example the Rindler coordinates in Minkowski space).
This should correspond to using the description of local observers who perceive divergent energy scale,
and can therefore have access to parts of the manifold, but not to the whole space-time~(as the Rindler observer cannot access the part of the manifold which is causally disconnected with him or her).
However, the global geometric and conformal invariants remain locally well defined. Since all dimensionless scalars are not changed by Weyl transformations, they remain well-defined
even in the case of singular conformal transformations. Note that $R$ or any observable $\mathcal{O}$ with conformal weight $w\neq 0$ do not fit in this category, but $R/\Phi^2$ and $\Phi^w\mathcal{O}$ (in $D=4$) do. Clearly all dimensionful 
quantities can become singular after a
singular conformal rescaling, but our assumption is that we cannot measure dimensionful parameters. 
Instead, we base our measurement on the local value of some field,
which just means that we measure $\Phi^w\mathcal{O}$ rather than $\mathcal{O}$. 
Conformal singularities might exist and they are point in which conformally invariant ratios diverge. However, they can always be 
mapped onto an infinite~(proper time) future or past.

\section{Coupling to matter}
\label{Coupling to matter}

In section~\ref{Conformal transformations in general relativity} we have showed that gravity with torsion, 
in the framework of Einstein-Cartan gravity, exhibits a geometrical version of conformal invariants.
Here we discuss how to construct a theory for scalar, spinor and vector fields, that exhibits the
same kind of conformal invariance in arbitrary space-time dimensions.

We start by defining a 1-form, given by the trace of the torsion tensor,
\begin{equation}
\label{torsion.form} \mathcal{T}\equiv \mathcal{T}_\mu \text{d}x^\mu = \frac{2}{D-1} T^\lambda{}_{\lambda\mu} \text{d}x^\mu \,.
\end{equation}
Note that the torsion trace is the only one of its irreducible components that transforms under
conformal transformations, and that the definition~(\ref{torsion.form}) is the unique metric independent
and coordinate independent contraction of the torsion tensor one can construct. We propose treating
the form~(\ref{torsion.form}) as the gauge boson of conformal transformations.
This choice is motivated by the fact that a conformal transformation changes $\mathcal{T}$ as,
\begin{equation}
\label{Conformal.Gauge.Symmetry} \mathcal{T} \rightarrow \mathcal{T} + \text{d}\theta \,,
\end{equation}
which is analogous to the way in which abelian gauge bosons transform, and by the fact that $\mathcal{T}$ acting on vectors
generates scale transformations, as a consequence of parallel transport.
The transformation law~(\ref{Conformal.Gauge.Symmetry}) has been noticed in the past, and has been tried
to be used to unify gravity with electromagnetism. For example, in Ref.~\cite{Maluf:1985fj} the author considers
the transformation law~(\ref{transformation.law.connection}) and the fact that the Riemann
tensor does not change upon applying it, and tries to link $\mathcal{T}$ to the gauge boson of $U(1)$.
However, even if the transformation law for~(\ref{Conformal.Gauge.Symmetry}) is identical to
the transformation law for the $U(1)$ connection, there is a key difference between the two: that
the Abelian group $U(1)$ is compact. On the contrary the conformal transformations that we
are studying in this paper form a non compact group, and is therefore to be distinguished from $U(1)$.
The parameter $\theta$ in~(\ref{Weyl.rescaling.0}) is non periodic, $\theta(x)\in (-\infty,+\infty)$, while in $U(1)$ transformations
one would write, for a field $\psi$, $\psi\rightarrow e^{iq\alpha(x)}\psi$, which shows that the space where the parameter $\alpha$ lives
requires the identification $\alpha \sim \alpha+ 2\pi$, {\it i.e.} it is a compact space.

Even though the concepts of $U(1)$ invariant derivative, and a conformally invariant derivative are distinct,
the way to construct them is analogous. We have already mentioned that scalar field in $D$ dimensions
transforms as,
\[\phi \rightarrow e^{-\frac{D-2}{2}\theta} \phi\,.\]
The conformally invariant covariant derivative can therefore be expressed as,
\begin{equation}
\label{Scalar.Field.Covariant.derivative} 
\overset{w}{\nabla}{}_\mu \phi = \partial_\mu \phi + \frac{D-2}{2} \mathcal{T}_\mu \phi 
=  \partial_\mu \phi + \frac{D-2}{D-1} T^\lambda{}_{\lambda\mu} \phi \,,
\end{equation}
and generalised to a field, $\boldsymbol{\Psi} $, of arbitrary conformal weight $w$ as, 
\begin{equation}
\label{Conformal.covariant.derivative.general.definition}\overset{w}{\nabla}{}_\mu \boldsymbol{\Psi}= \nabla_\mu \boldsymbol{\Psi} + (w_g - w) \mathcal{T}_\mu \boldsymbol{\Psi}\,,
\end{equation}
where $\nabla_\mu$ is the manifold covariant derivative, and $w_g$ is the geometrical dimension of $\boldsymbol{\Psi}$, 
that is if $\boldsymbol{\Psi}$ is a $\binom{p}{q}$ tensor, $w_g = q-p$. Note that in order to be able to construct $\overset{w}{\nabla}$ for a given field, 
we should know its scaling dimension, $w$. This is not different from the gauge derivative of fields charged under $U(1)$: in that case, one should know the 
hypercharge of the representation upon which the gauge derivative acts, $Y$, which is different for different fields. The role of hypercharge is played, for the conformal group, by the scaling dimension of fields, $w$. 

We can think of the field $\boldsymbol{\Psi}$ as being a representation 
of the conformal extension of the Lorentz symmetry group, which is classified by its conformal weight $w$. Clearly $\boldsymbol{\Psi}$ is going to be also a representation 
of the Lorentz group, which will give it a ``natural'' conformal weight: under Lorentz transformations $\boldsymbol{\Psi} \rightarrow \boldsymbol{\Lambda}  \boldsymbol{\Lambda} \cdots  \boldsymbol{\Lambda}^{-1}  \boldsymbol{\Lambda}^{-1} \cdots\boldsymbol{\Psi}$, 
where there are $q$ $ \boldsymbol{\Lambda}$'s and $p$ $ \boldsymbol{\Lambda}^{-1}$'s. Under global scale transformations we have, 
\[x\rightarrow \lambda x \implies \boldsymbol{\Psi} \rightarrow (\mathds{1}\lambda) (\mathds{1}\lambda) \cdots  (\mathds{1}\lambda^{-1}) (\mathds{1}\lambda^{-1})\cdots\boldsymbol{\Psi} = \lambda^{(q-p)}\boldsymbol{\Psi}\, ,\]
which sets its ``Lorentz'' or geometrical weight to $q-p$, when the global scaling behaviour is made local, {\it i.e.} $\lambda\rightarrow\lambda(x)$. We can however form composite objects out of Lorentz scalars with $w\neq 0$, and representations $\boldsymbol{\Psi}$ 
having $w=w_g$. One example of such field constructed using geometrical quantities is $\Theta^w \dot{\gamma}{}^\mu$, a vector with conformal weight $-w-1$. 
This shows that the conformal weight of fields can, in general, take any real value, which is a consequence of the non-compactness of the conformal group. Thinking back again to the 
$U(1)$ example: the electric charge is quantised because of the global identification $\alpha \sim \alpha + 2\pi$~\cite{Frankel:1997ec}. This is not the case 
for the conformal group, whose representations can therefore possess any scaling behaviour.
If $w$ is an integer, $w-w_g$ simply represents the energy dimension, in natural units, of the field $\boldsymbol{\Psi}$. Fields for which $w=w_g$ are dimensionless in natural units, 
as for example $\dot{\gamma}{}^\mu =\text{d}x^\mu /\text{d}\tau $, measured in units of [space]/[time] and as such dimensionless in natural units. 

By following this procedure, one can construct the covariant conformal derivative for spinor fields and vector bosons 
using the transformation laws,
\begin{eqnarray}
\label{canonical.scaling.fermions}\psi \rightarrow e^{-\frac{D-1}{2}\theta} \psi \,,\\
\label{canonical.scaling.bosons} A_{\mu} \rightarrow e^{-\frac{D-4}{2}\theta} A_\mu \,.
\end{eqnarray}
The form of $\overset{w}{\nabla}{}$ for gauge fields follows from Eq.~(\ref{Conformal.covariant.derivative.general.definition}), and for fermions we define,
\begin{eqnarray}
\label{conformal.derivative.fermions}
\overset{w}{\nabla}{}_\mu \psi &=& {\nabla}_\mu \psi +\frac{D-1}{2} \mathcal{T}_\mu\psi
 =  {\nabla}_\mu \psi + T^\lambda{}_{\lambda\mu}\psi  \,,\\
\label{conformal.derivative.bosons} 
\overset{w}{\nabla}{}_\mu A_\nu &= &\nabla_\mu A_\nu + \frac{D-2}{2} \mathcal{T}_\mu A_\nu = \overset{\circ}{\nabla}_\mu A_\nu + \mathcal{T}_\nu A_\mu - g_{\mu\nu} \mathcal{T}_\sigma A^\sigma +\frac{D-2}{2}  \mathcal{T}_\mu A_\nu  \,,
\end{eqnarray}
where ${\nabla}_\mu$ is the covariant derivative with torsion satisfying metric compatibility, $\overset{\circ}{\nabla}_\mu$ the part of the covariant derivative that depends on the metric only. 
Note that $\overset{w}{\nabla}{}_\alpha g_{\mu\nu} = 0$, since the conformal weight of $g_{\mu\nu}$ coincides with its 
geometrical weight. Furthermore, it is easy to check that $\overset{w}{\nabla}{}_\mu$ satisfies the Leibniz rule, and 
commutes with contractions and tensor product. It is also conformally and coordinate invariant~\footnote{There exist other choices to construct a conformally invariant derivative, namely, using $\Gamma^\lambda{}_{\lambda\mu}/D$ or $\Gamma^\lambda{}_{\mu\lambda}$ in place of $\mathcal{T}_\mu$. However, such choices are not covariant with respect to Lorentz transformations, because the Christoffel symbols do not transform as tensors.}.

We point out that the construction that lead to Eq.~(\ref{Conformal.covariant.derivative.general.definition}) is not guaranteed to be unique. 
This means that there could be other definitions of covariant derivatives that are conformal, coordinate invariant and metric preserving, and that Eq.~(\ref{Conformal.covariant.derivative.general.definition}) might not
be the correct generalisation of the space-time covariant derivative. 
Nevertheless Eq.~(\ref{Conformal.covariant.derivative.general.definition}) defines a derivative operator on the manifold $\mathcal{M}$,
which satisfies the basic properties of derivations and is coordinate and conformally invariant, for a field of arbitrary conformal weight $w$. 
Furthermore, Eq.~(\ref{Conformal.covariant.derivative.general.definition}) reduces to the usual covariant derivative of space-time, when it acts on 
a field of energy dimension $0$, that is when $w=w_g$. We therefore consider it appropriate for the time being, and we will proceed, in next section, to write
conformally invariant actions for scalars, fermions and gauge bosons. We will limit our discussion
to classical theories and postpone any consideration on the quantum behaviour of the theory to forthcoming
publications.

\subsection{Scalars}
\label{Scalars}

We can clearly write the kinetic term for the scalar field with internal symmetry group $G$ as,
\begin{eqnarray}
\label{kinetic.term.scalar} 
&&-\frac12\int \text{d}^D x \sqrt{-g}{\rm Tr} \overset{w}{\nabla}{}_\mu \phi \overset{w}{\nabla}{}_\nu \phi g^{\mu\nu} =
\nonumber\\
&&\hskip 1cm = -\frac12\int \text{d}^D x \sqrt{-g}{\rm Tr} \left (\partial_\mu \phi +\frac{D-2}{2} \mathcal{T}_{\mu} \phi  \right ) 
      \left (\partial_\nu \phi +\frac{D-2}{2} \mathcal{T}_{\nu} \phi  \right )g^{\mu\nu} 
\,,
\end{eqnarray}
which is invariant under conformal transformations. 
Here $\phi = \sum\limits_a \phi^a \lambda^a$, where $\lambda^a$ are the generators of the group $G$ of internal symmetries
and the trace ${\rm Tr}$ acts in the internal group space (for a real scalar field ${\rm Tr}$ is a trivial operation). 
Because of this we can write the following interaction terms 
\begin{equation}
\label{scalar.field.interaction} \int\text{d}^D x \sqrt{-g} {\rm Tr}\left ( \frac{ \phi^2}{2\alpha^2} R - \lambda \phi^4  \right )\,.
\end{equation}
where $\alpha$ and $\lambda$ are coupling constants. Note that, while the first term is conformally invariant in general $D$, 
the second is only in $D=4$. This also means that $\alpha$ is dimensionless in general $D$, while $\lambda$ is dimensionless only 
in $D=4$ (the canonical dimension of $\lambda$ is $-(D-4)$).

In $4$ space-time dimensions operators of dimension $4$ in the pure gravity sector can be added to
our theory without spoiling conformal invariance, and they are generic in the sense that they are always generated 
by quantum fluctuations~\cite{birrell1984quantum}. For example, the following effective action,
\begin{equation}
\label{Gravitational.Lagrangian} \int \text{d}^Dx \sqrt{-g} \left ( \xi_1 R^2 + \xi_2 R_{\mu\nu} R^{\mu\nu} + \xi_3 R_{\alpha\beta\gamma\delta}R^{\alpha\beta\gamma\delta}\right ) \,,
\end{equation}
emerges generically when the (one-loop) quantum corrections of scalars, vectors and fermions 
are taken account of and it is conformally invariant 
in four dimensions (in the sense discussed in this paper).

Note further that the space of conformally invariant theories that one can construct using
conformal symmetry in Einstein-Cartan gravity is much wider and much less constrained
than conformally invariant theories containing the metric alone. There the only choice we have is
to write the square of the Weyl tensor, or choose the non minimal coupling $1/\alpha^2$ between the scalar
field and the Ricci curvature to be $(D-2)/[4(D-1)]$.

It is worth spending a few words to analyse what theory emerges for the Higgs particle, and how its effective
low energy description can reproduce the Higgs action of the standard model. Writing $H = \sum\limits_{a=0}^3 H^a \sigma^a/2$,
where $\sigma^a/2$ are the $SU(2)$ group generators~($\sigma^a$ are here the Pauli matrices, and $\sigma^0=\mathds{1}$ is the group identity element),
we can write the Higgs conformal-gauge derivative as,
\begin{equation}
\label{Higgs_Full_Covariant_Derivative}
  D{}_\mu H = \partial_\mu H + \frac{D-2}{2} \mathcal{T}_\mu H -ig \sum\limits_{a}  W^a_\mu \sigma^a \cdot H-ig'Y B_\mu H  \,,
\end{equation}
where $\sigma^a \cdot H$ denotes the product in the $SU(2)$ group space, 
$g$ is the weak gauge coupling constant, $g^\prime$ is the hypercharge gauge coupling constant,
$Y=1$ is the hypercharge of the Higgs doublet and 
$B_\mu$ is the (Abelian) hypercharge field.
The action for the Higgs field then gets modified to,
\begin{equation}
\label{Higgs_Action}
\int\text{d}^Dx \sqrt{-g}\left [-\frac{1}{2}(D{}_\mu H)^\dagger D{}^\mu H - \lambda_H (H^\dagger H)^2 + g_{H\Phi} H^\dagger H\Phi^2-\lambda_\Phi \Phi^4
\right]
\,,
\end{equation}
where, in order to make the action conformally invariant in $D=4$, we traded the Higgs mass for a dilaton field $\Phi$.
This theory can exhibit spontaneous symmetry breaking in the following sense: 
at high energies both $\langle H\rangle$ and $\langle\Phi\rangle$ are close to
zero, as required by conformal symmetry. When the energy scale drops below a critical value, $\langle\Phi\rangle$ starts growing 
towards a finite value~(possibly driven by the non-minimal coupling $\Phi^2 R$). 
This process will make the Higgs potential develop a new non trivial minimum, and as a consequence
also the Higgs field will develop a vacuum expectation value.

When the coupling constants satisfy, $\lambda_H \lambda_\Phi = g_{H\Phi}^2/4$, then the effective low energy action becomes,
\begin{equation}
\label{Higgs_Action}\int\text{d}^Dx \sqrt{-g}\left[-\frac{1}{2} (D{}_\mu H)^\dagger D{}^\mu H - \lambda_H \left (H^\dagger H - \frac{g}{2\lambda_H} \langle\Phi^2\rangle\right )^2\right]
\,,
\end{equation}
which leads to the Higgs {\it vev},
 $2\langle H^\dagger H\rangle = h^2 = \frac{g}{\lambda_H} \langle\Phi^2\rangle=(246~{\rm GeV})^2$. 
Note that this value of the Higgs {\it vev} produces
a classical cosmological constant exactly equal to zero, {\it i.e.} $\Lambda \propto  \left (\langle H^\dagger H\rangle - \frac{g}{2\lambda_H} \langle\Phi^2\rangle\right )^2 = 0$.
Of course this does not take into account quantum effects, which can produce a non vanishing cosmological constant, 
also by making $\lambda_H, \lambda_\Phi, g_{H\Phi}$ run with the energy scale.

In Ref.~\cite{GarciaBellido:2011de}
 the authors consider a model similar to the one defined by~(\ref{Higgs_Action}), and show that it possesses inflationary solutions
and exhibit late time dark energy domination. The model proposed in~\cite{GarciaBellido:2011de} exhibits a global scale symmetry, while the one that we are proposing in this
paper makes the symmetry local by introducing a coupling between the scalar fields and torsion. If we conjecture that local conformal symmetry should be realised at high energy,
the theory proposed in this paper can be seen as the UV completion of the model in~\cite{GarciaBellido:2011de}, which could in principle explain inflation and late time dark energy,
while providing an interesting framework to study the microscopic properties of gravity with torsion.

\subsection{Fermions}
\label{Fermions}

It is well known that in general relativity and in general space-time dimension 
the kinetic term of a fermionic field can be written as,
\begin{equation}
\label{kinetic.term.fermions} 
\int \text{d}^D x \sqrt{-g} 
\frac{i}{2} \left[ \bar{\psi} \gamma^\mu {\nabla}_\mu \psi 
   -  \big({\nabla}_\mu \bar{\psi}\big) \gamma^\mu \psi \right]
\,,
\end{equation}
where
\begin{eqnarray}
\label{cartan.derivative.spinor}{\nabla}_\mu \psi &=& \partial_\mu \psi - \frac{1}{8}\omega^{ab}{}_\mu [\gamma_a,\gamma_b] \psi  \,, \\
{\nabla}_\mu \bar{\psi} &=&\partial_\mu\bar{\psi} + \frac{1}{8}\omega^{ab}{}_\mu \bar{\psi} [\gamma_a,\gamma_b] \,.
\end{eqnarray}
Here $\omega^{ab}{}_\mu$ is the spin connection defined by,
\begin{equation}
\label{spin_connection} \omega^{ab}{}_\mu = e^a_\lambda\left ( \partial_\mu e^{b\lambda} + \Gamma^\lambda{}_{\sigma\mu} e^{\sigma b}\right )\,
\end{equation}
and $\bar{\psi}$ is defined by
\[\bar{\psi} = \psi^\dagger \tilde{\gamma} \,,\]
where $\tilde{\gamma}$ satisfies,
\[\left (\gamma^\mu\right )^{\dagger} = \tilde{\gamma}\gamma^\mu\tilde{\gamma}\,,\]
and it is therefore invariant under conformal transformations. This can be shown
by using the definition of the $\gamma^\mu$ matrices, to find their scaling properties under conformal transformations,
\[\{\gamma_\mu, \gamma_\nu\} = g_{\mu\nu} \implies \gamma^\mu \rightarrow e^{-\theta} \gamma^\mu \,,\]
and using the fact that the gauge parameter $\theta$ is real, we find that $\bar{\psi}$ transforms like $\psi$.

It would be tempting to argue that the vierbein field $e_a^\mu$
transforms in the same way as $\gamma^\mu = e^\mu_a \gamma^a$, however this would be incompatible with the assumption that the tangent space metric is parallel transported, {\it i.e.} $\nabla_\mu \eta_{ab}=0$.
In fact in the Cartan formalism it is the flat metric that transforms under conformal transformations, 
which is the only way in which the flat metric remains parallel with respect to the new connection. 
Indeed, if the tetrad does not transform
we have,
\begin{equation}
\omega^a_{b\mu} \rightarrow \omega^a_{b\mu} + \delta^a_b \partial_\mu \theta
\,,
\label{spin connection transformation}
\end{equation}
under conformal rescaling, which immediately implies that,
\[\nabla_\mu \eta_{ab} = 0  \rightarrow\nabla_\mu \tilde{\eta}_{ab} -2 \partial_\mu \theta \,\tilde{\eta}_{ab}=0 \implies \tilde{\eta}_{ab} = e^{2\theta} \eta_{ab}\,.
\]
We speculate that the reason for this is that the manifolds that we are considering are not locally isomorphic to flat spaces, 
but instead to {\it conformally} flat spaces.

In light of this comment, we notice that we can rewrite the fermionic covariant derivative as, 
\begin{equation}
\label{Fundamental.Covariant.Derivative} \nabla_\mu \psi =  \partial_\mu \psi - \frac{1}{8}\omega^{[ab]}{}_\mu \big[\gamma_a,\gamma_b\big]\psi - \frac{1}{8}\omega^{(ab)}{}_\mu \big\{\gamma_a,\gamma_b\big\}\psi\,,
\end{equation}
which will lead to Eq.~(\ref{cartan.derivative.spinor}) in the general relativity gauge, that is where the connection $\omega^{ab}{}_\mu=0$ is anti symmetric in $(a\leftrightarrow b)$ and the covariantly conserved metric is $\eta_{ab}$. 
Evaluating for the connection $\omega^{ab}{}_\mu$ as in~(\ref{spin connection transformation}), we get,
\begin{equation}
\nabla_\mu \psi =  \partial_\mu \psi - \frac{1}{8}\omega^{[ab]}{}_\mu \big[\gamma_a,\gamma_b\big]\psi - \frac{D}{4}\partial_\mu \theta \psi\,,
\end{equation}
which is conformally invariant, if the conformal weight of $\psi$ is $w_\psi = D/4$, which we can call the geometrical weight of spinor fields~(in $D=4$, $w_\psi =1$, since under Lorentz transformations $\psi \rightarrow \boldsymbol{\Lambda} \psi$).  
Notice that this derivative splits into a part proportional to the spinorial generators of Lorentz transformation, $ \big[\gamma_a,\gamma_b\big]$, 
and a part proportional to the spinorial generators of conformal transformations, $\mathds{1}$.

In view of~(\ref{spin connection transformation}), the kinetic term~(\ref{kinetic.term.fermions}) is 
conformal in any number of dimension. 
However, the coupling of fermions to gauge fields can be made conformal only in four dimensions since,
\[\sqrt{-g}\bar{\psi} \gamma^\mu A_\mu \psi \rightarrow \sqrt{-g} e^{-\frac{D-4}{2} \theta}\bar{\psi} \gamma^\mu A_\mu \psi \,.\]
Also the Yukawa couplings are conformal only in four dimensions since,
\begin{equation}
\sqrt{-g}\phi \bar{\psi}\psi \rightarrow e^{-\frac{D-4}{2}\theta} \sqrt{-g}\phi \bar{\psi}\psi \,.
\end{equation}
Hence we see that all couplings between fermions and gauge bosons or scalar fields in the Standard Model  
are conformal in four dimensions. In four dimensions the Standard Model Lagrangian for fermions
is therefore,
\begin{equation}
\label{Conformal.Fermions} 
\int \text{d}^4 x \sqrt{-g} \left [ \frac{i}{2}\left ( \bar{\psi} \gamma^\mu( \nabla_\mu+e A_\mu) \psi - ( \nabla_\mu-eA_\mu) \bar{\psi} \gamma^\mu \psi \right ) - g_y \phi \bar{\psi}\psi\right ]\,,
\end{equation}
and with $\nabla_\mu = \stackrel{\circ}{\nabla}_\mu$ this action is invariant both under conformal and gauge transformations.
The action~(\ref{Conformal.Fermions}) can be easily generalized to non-Abelian groups $G$ by writing 
$A_\mu=A_\mu^a \lambda^a$, where $\lambda^a$ the suitable generators of the group and a trace is taken over the group $G$.
A similar generalization of scalar and fermionic fields is in order, as can be found in any textbook on the Standard Model.

\subsection{Gauge fields}
\label{Gauge fields}

As we have remarked before, gauge fields do not transform under conformal transformations
only in four dimensions. It is in fact worth noticing that $D=4$ is the only dimension in which the
gauge field lagrangian can be made invariant simultaneously under conformal transformations and gauge 
transformations~\footnote{This fact alone is of some interest, as it could be used as a starting point for 
the explanation of why we live in a four dimensional space-time.
}.
In fact, in general dimensions the conformally invariant field strength has to be written as,
\begin{equation}
\label{Conformal.field.strength} F^a_{\mu\nu} =\overset{w}{\nabla}{}_{[\mu} A^a_{\nu]} =\partial_{[\mu} A^a_{\nu]} + \frac{D-4}{2} \mathcal{T}_{[\mu} A^a_{\nu]} \,,
\end{equation}
and therefore spoils gauge invariance. Conversely, if we want to write a field strength that preserves gauge symmetry,
we have to neglect the second term in Eq.~(\ref{Conformal.field.strength}), which will spoil conformal symmetry.

Luckily we live in four dimensions, where the gauge field action,
\begin{equation}
\label{Gauge.field.invariant.action}-\frac{1}{4} \int \text{d}^4 x \sqrt{-g} \text{Tr} \left ( F_{\mu\nu}F^{\mu\nu}\right ) \,,
\end{equation}
is both conformally and gauge invariant. Similarly, one can show that the action of the form,
\begin{equation}
 \int d^Dx f {\rm Tr}\left[F_{\mu\nu}\tilde F^{\mu\nu}\right]
 \nonumber
 \,,
\end{equation}
which is conformal (and topological) in $D=4$, but not conformal (non-topological) in $D\neq 4$.
Here $f$ is a coupling constant, $\tilde F^{\mu\nu}=|g|^{-1/2}\epsilon^{\mu\nu\alpha\beta}F_{\alpha\beta}$ 
is the dual field strength, which under conformal transformations
transforms the same way as $F^{\mu\nu}$,  and $\epsilon^{\mu\nu\alpha\beta}$ is the Levi-Civit\`a symbol.

\subsection{Gravity plus dilaton: a toy model}
\label{Gravity plus dilaton: a toy model}

To end this section, we are going now to solve a toy version of this model in the classical limit. Namely, we assume that
there is only one real scalar field, $\Phi$, that couples non minimally to gravity and therefore sets the Planck scale.
Clearly this is not the realistic situation, since there should be at least one extra scalar in the model, the Higgs field,
and it is charged under $SU(2)$. However, $\Phi^2 = \sum_a(\phi^a)^\dagger \phi^a$, 
can be thought of as an effective sum of all scalars that non minimally couple to gravity,
and at the classical level our model can therefore be realistic. For simplicity, we also assume
that the only non vanishing part of torsion is its trace. Since at the level of the Ricci scalar the torsion trace
and the skew symmetric part of torsion decouple, since fermions only source the skew symmetric part of torsion~\cite{Lucat:2015rla}, and since the remaining irreducible part of torsion is not sourced
by any matter, our considerations are general. The action then reads,
\begin{equation}
\label{Action.Toy.Model} 
S[g_{\mu\nu}, \Phi, \mathcal{T}_\mu] = \int \text{d}^4 x \sqrt{-g} \left (\frac{ \Phi^2}{2\alpha^2} R 
- \frac{g^{\mu\nu}}{2} \overset{w}{\nabla}{}_\mu\Phi \overset{w}{\nabla}{}_\nu \Phi - V(\Phi)\right ) +S^{\it SM}
 \,,
\end{equation}
where $S^{\it SM}$ is the action of fermions and gauge fields which, in four dimensions, does not depend on the torsion trace.

Varying the action with respect to $\mathcal{T}_\nu$ and $g^{\mu\nu}$ leads to the classical equations of motion, 
which read~\footnote{Note that the usual choice of conformally coupled scalar, 
$\alpha^2=
6$ in $D=4$ here leads to no constrain on torsion. 
This is so because that specific choice of $\alpha^2$ leads to cancellation of all the torsion contributions in the action~(\ref{Action.Toy.Model}).},
\begin{eqnarray}
\label{scalar.field.torsion.equations}\overset{w}{\nabla}{}_\sigma \overset{w}{\nabla}{}^{\sigma} \Phi = \left (\overset{\circ}{\nabla}{}_\sigma -\mathcal{T}_\sigma\right )\overset{w}{\nabla}{}^{\sigma} \Phi &=& -\frac{1}{\alpha^2} R \Phi + \lambda \Phi^3\\ 
\label{Maxwell.Torsion.Equations.1}
   (6-\alpha^2)  \Phi \overset{w}{\nabla}{}_\nu \Phi=\frac{(6-\alpha^2)}{2}\overset{w}{\nabla}{}_\nu \Phi^2  &=& 0 \,, \\
\label{Einstein.Torsion.Equations}\hskip -0.5cm 
R_{\mu\nu} -\frac{1}{2} g_{\mu\nu} R - \frac{1}{\Phi^2}\bigg [\left (\overset{\circ}{\nabla}{}_\mu 
+ 4\mathcal{T}_\mu\right ) \overset{w}{\nabla}{}_\nu \Phi^2
&-&  g_{\mu\nu} \left (\overset{\circ}{\nabla}{}_\sigma+\mathcal{T}_\sigma\right) \overset{w}{\nabla}{}^{\sigma} \Phi^2\bigg ] 
= \frac{\alpha^2}{\Phi^2} T^{\it m}_{\mu\nu} 
\,,
\end{eqnarray}
where $\overset{\circ}{\nabla}{}_\mu$ is the covariant derivative computed using the metric and the Christoffel symbols,
and $\overset{w}{\nabla}$ is the conformal covariant derivative from Eq.~(\ref{Scalar.Field.Covariant.derivative}).

For our toy model, the matter stress energy tensor is going to be,
\begin{equation}
\label{Stress.Energy.Tensor.Toy.Model}T^{\it m}_{\mu\nu} 
=\overset{w}{\nabla}{}_\mu\Phi \overset{w}{\nabla}{}_\nu\Phi   
- g_{\mu\nu} \left (\frac{1}{2}\overset{w}{\nabla}{}_\sigma\Phi \overset{w}{\nabla}{}^\sigma\Phi + V(\Phi)\right ) + T^{\it SM}_{\mu\nu}\,,
\end{equation}
where now $T^{\it SM}_{\mu\nu}$ is the energy-momentum tensor fermions and gauge 
fields~\footnote{In this toy model the Higgs contribution may (but need not) be absorbed into $\Phi$.}.

The non trivial solution of Eq.~(\ref{Maxwell.Torsion.Equations.1}) is,
\begin{equation}
\label{Intrinsic.Scale.Definition} 
\mathcal{T}_\mu(x) = -\frac{1}{2}\partial_\mu \log \frac{\Phi^2(x)}{\Phi_0^2} \,,
\end{equation}
where we introduced the~(arbitrary) scale $\Phi_0^2$, to make the argument of the logarithm dimensionless. It represents 
an arbitrary energy scale, {\it i.e.} the value of $\Phi(x_0)$ at some arbitrary point $x_0$, such that the ratio $\Phi^2(x)/\Phi_0^2$ 
measures the variation of the field.
We have thus arrived to an equation, valid in the classical limit of the theory, which shows the connection between the
intrinsic scale that an observer uses to measure its distances and the transformation law~(\ref{Weyl.rescaling.0}).
Note that Eq.~(\ref{Intrinsic.Scale.Definition}) implies that the scalar field is covariantly conformally conserved, or that,
\begin{equation}
\label{parallel.transport.intrinsic.scale} \overset{w}{\nabla}{}_\mu \Phi(x) = \frac{1}{2\Phi(x)}\overset{w}{\nabla}{}_\mu \Phi^2(x) = 0 \,.
\end{equation}
If Eq.~(\ref{Intrinsic.Scale.Definition}) is valid, it means that the torsion is in its pure gauge form from Eq.~(\ref{transformation.law.connection}),
which in turn implies that the metric has to be in the form $\Phi^2 \otimes\hat{g}$, where $\hat{g}$ is the metric in the General Relativity
gauge~({\it i.e.} where the torsion trace vanishes). Because of Eq.~(\ref{Intrinsic.Scale.Definition}), we can argue that the metric which is parallel transported has to be,
\begin{equation}
\label{Local.Observers.Scale} 
\text{d}s^2 = \frac{\Phi_0^2}{\Phi^2(x)} \text{d}\hat{s}^2 = \frac{\Phi_0^2}{\Phi^2(x)} \hat{g}_{\mu\nu} \text{d}x^\mu\text{d}x^\nu \,,
\end{equation}
where $\Phi_0$ is an integration constant with 
the dimension of energy and $\hat{g}_{\mu\nu}$ solves the
effective equation,
\begin{equation}
\label{Einstein.Effective} \overset{\circ}{R}{}_{\mu\nu}[\hat{g}] - \frac{1}{2} \hat{g}_{\mu\nu} \overset{\circ}{R}[\hat{g}] = \frac{\alpha^2}{\Phi_0^2} \hat{T}{}^{\it SM}_{\mu\nu} -\alpha^2 \hat{g}_{\mu\nu}\Phi_0^2 \left (\frac{V(\Phi)}{\Phi^4}\right )\,,
\end{equation}
where $\circ$ denotes as usual quantities computed using only the metric without torsion, in this case the metric $ \hat{g}_{\mu\nu}$ in Eq.~(\ref{Local.Observers.Scale}), and 
\[\hat{T}{}^{\it SM}_{\mu\nu} = -\frac{2}{\sqrt{-\hat{g}}} \frac{\delta S^{\it SM}}{\delta \hat{g}{}^{\mu\nu}}\,.\]
Now, we note that $\Phi$ is not a dynamical field, in fact if $\overset{w}{\nabla}{}_\mu \Phi = 0$, Eq.~(\ref{scalar.field.torsion.equations})
turns non dynamical, and it is solved by $\Phi = 0$ and, if $R> 0$, by $\Phi^2 = R/{\lambda\alpha^2}$. Plugging this in Eq.~(\ref{Einstein.Effective}), 
leads to 
\begin{equation}
\label{Einstein.Effective} \overset{\circ}{R}{}_{\mu\nu}[\hat{g}] - \frac{1}{2} \hat{g}_{\mu\nu} \overset{\circ}{R}[\hat{g}] = \frac{\alpha^2}{\Phi_0^2} \hat{T}{}^{\it SM}_{\mu\nu} -\lambda \alpha^2 \hat{g}_{\mu\nu}\Phi_0^2 \,,
\end{equation}
which are Einstein's equations with a positive cosmological constant. We notice that such a solution only exists if $R> 0$, that is in our notation de Sitter space~\footnote{Note that the de Sitter metric would be an exact solution, for the metric $\hat{g}_{\mu\nu}$, of Eq.~(\ref{Einstein.Effective}), if 
$\hat{T}{}^{\it SM}_{\mu\nu}= 0$.}, which is also supported by the results in~\cite{Glavan:2015ora} where the authors find that a condensation of the scalar field is only possible in de Sitter space-time. However, in our theory, the restriction $R>0$ only holds when 
there is only one dilaton field $\Phi$. If more scalars are introduced, they would all turn dynamical and the space of solutions will become bigger, and not restricted to $R>0$. Even in this situation 
we can solve exactly for the torsion trace, since Eq.~(\ref{Maxwell.Torsion.Equations.1}) would still not contain kinetic terms for torsion. Then, a cosmological constant $\Lambda_{\it eff} = \lambda \alpha^2\Phi_0^2$ 
would still be generated. We want to estimate the magnitude of such $\Lambda_{\it eff}$: it is clear that the fraction $\frac{\Phi_0^2}{\alpha^2}$ has to be of order $1$ in Planck units, because it gives the measured value of the Planck mass. 
However, this ratio does not change if we multiply both $\alpha$ and $\Phi_0$ by the same dimensionless constant $C$. The cosmological constant, however, is going to change by a factor $C^4$, which, for 
$C\ll 1$, can render it arbitrarily small. This argument shows that a small cosmological constant can be realised in a natural way in this theory, also considering that we can vary $\lambda$ arbitrarily.

The metric~(\ref{Local.Observers.Scale}) clearly splits in two different representations of the symmetry group defined by~(\ref{Weyl.rescaling.0}): 
conformal transformations change $\Phi^{-2}$, in~(\ref{Local.Observers.Scale}), while Lorentz transformations 
only act on $\Phi_0^2\text{d}\hat{s}^2 = \Phi_0^2\hat{g}{}_{\mu\nu}\text{d}x^\mu\text{d}x^\nu$. From this point of view, 
the metric is a composite object that contains two very distinct parts: a dimension-full part, $\Phi^{-2}$, is related to the Planck mass, 
while the dimensionless part, $\Phi_0^2\hat{g}{}_{\mu\nu}\text{d}x^\mu\text{d}x^\nu$, is the metric that solves Einstein's equations. 

Eq.~(\ref{Local.Observers.Scale}) sets the form of the metric that an observer uses to measure proper time.
Now let us consider an observer performing local experiments: any measure he performs will be compared with
the only local scale he observes, that is the Planck mass. However, since the Planck mass in our theory is given
by the field $\Phi$~(see Eq.~(\ref{Conformal.Einstein.Equations})), the local scale of observers is set by the field $\Phi$ itself. Note that this is precisely the
interpretation of Eq.~(\ref{Local.Observers.Scale}): modulo a constant proportionality factor, the natural length unit is
set in this theory by the Planck scale. This is in line with our comment in the introduction: the transformation law~(\ref{Weyl.rescaling.0})
is really just a change of reference frame, switching from different observers that perceive locally different physical scales.
Eq.~(\ref{Local.Observers.Scale}) then shows that the natural scale to measure proper lengths is set by the dilaton which
produces the Planck scale.

The interpretation of space-time singularities differs in this theory from the mainstream interpretation: from the form of~(\ref{Local.Observers.Scale}) and~(\ref{Einstein.Torsion.Equations})
one can infer that singularities are points where $\Phi= 0$. Since $\text{d}s^2$ diverges when this situation is 
realised~(as one infers from~(\ref{Local.Observers.Scale})), it would take an infinite amount of proper time
to reach such singular points. This in particular means that collapsing matter can never reach the singularity. Since physical black holes eventually evaporate, all the matter that has fallen into it
will be released, once the horizon shrinks enough.

In this process, the physical separation of a congruence of geodesics, {\it i.e.} $\|J_\perp\| =\|\hat{J}{}_\perp\|/\Phi^2$ defined as in section~\ref{Conformal transformations in general relativity},
might not go to zero, even if $\|\hat{J}{}_\perp\|$~(the separation in Einstein's frame) does. When conjugate points form we have $\|J{}_\perp\|\rightarrow 0$, which can happen only in the asymptotic proper-time future, and might
be even prevented by the dynamics of the field $\Phi$.

\section{Conclusion}
\label{Conclusion}

In this paper we show that it is possible to formulate a conformally invariant theory of gravity
in the context of Einstein-Cartan gravity. At the classical level and if torsion is in the pure gauge form given by equation~(\ref{transformation.law.connection})
such a theory reproduces general relativity with an extra symmetry .
Its deviation from Einstein's theory manifests itself in two ways: firstly, the symmetry demands a field-dependent
Planck constant which can, for example, be described by a (dynamical) scalar field, and secondly by the
the presence of torsion. According to our considerations on cosmological observations in the introduction, and because torsion
contributions are only noticeable at very high energies~\cite{RevModPhys.48.393}, both of these possibilities
are not ruled out by current observations.

In section~\ref{Coupling to matter} we construct a classical conformally invariant theory which can be extended to include gravity
and all fields of the Standard Model. We show that gauge invariance and conformal invariance are only compatible in four dimensions, another surprising coincidence which
motivates further study of the class of models presented in this paper. In our theory the torsion trace 
is the ``gauge boson" of the conformal group and, as we show in
Eq.~(\ref{Intrinsic.Scale.Definition}), in the classical limit of the theory it collapses to its pure gauge form and selects the scale used
by local observers to measure lengths.

These considerations show a novel connection, guided by symmetry, between the standard model of particle physics and gravity, which needs to
be understood in full. Conformal symmetry might in fact be an enhanced symmetry of gravity, 
which reflects the fact that different observers might measure locally different
scales, but have to agree on the~(dimensionless) results of their experiments. Since the classical theory we formulate in this paper
is exactly conformally invariant, we are confident that it is the right road to build models with this extended equivalence principle.

In this paper, we have not studied any of the quantum properties of the theory, which is of course of fundamental importance in the future.
It is a well established fact that in general quantum fluctuations do not respect conformal symmetry~\cite{Meissner:2006zh}, since the path integral measure is not invariant under conformal rescaling of the fields. 
However, our theory is power-counting  renormalizable, since it only contains dimensionless
coupling constants, and should therefore posses an UV fixed point of the renormalization group. 
What is important to notice is that, near the fixed point,
quantum fluctuations restore conformal symmetry due to the very nature of the fixed point. Therefore at the UV fixed point of the
theory the symmetry is restored. If the evolution of the universe starts close to this UV fixed point, the conformal symmetry might be mildly broken,
eventually leading to a condensation of the scalar fields, which would then be responsible for the generation of mass. 
One hopes that by studying these processes in detail one can arrive at satisfactory inflationary dynamics that result in predictions consistent with the observations.
Our mechanism for inflation would be similar to that studied in~\cite{GarciaBellido:2011de}, where the field responsible for generating the Planck mass is also responsible for inflation. The crucial difference is, however, in the gravitational sector which in our model 
exhibits torsion. Since gravitational dynamics is relevant for inflationary predictions, 
it would be of crucial importance to find out what are 
the observable that would differentiate between the model studied in~\cite{GarciaBellido:2011de} and our model.

It would also be interesting to ask the question whether our theory can be obtained by a low energy description of string theory: compactification procedures usually are expected to break
conformal symmetry, as there is always a scale involved in such procedures, {\it i.e.} the radius of the compactified dimensions. 
Therefore, the breakdown of conformal symmetry in low energy models that arise from string theory compactifications could be 
traced back to the scale set by compactification.
Alternatively, local conformal symmetry can emerge from embedding 4-branes in a higher dimensional space-time, 
and its realisation be confined to the brane itself.

Last but not least, other open questions linked to the quantum behaviour of the theory are related to (quantum) anomalies. 
Since conformal anomalies are produced, in dimensional regularization scheme,
by a ``memory'' that the fields seem to have of extra dimensions~\cite{birrell1984quantum}, and since conformal and gauge invariance are only compatible in four dimensions in our theory,
there is the potential for gauge anomalies. This question ought to be answered by carefully studying 
the quantum behaviour of the theory.

 \section*{Acknowledgements}

The authors would like to thank Anton Quelle and Huibert Het Lam for useful discussions about the Jacobi equation. 
This work is part of the D-ITP consortium, a program of the Netherlands Organisation for Scientific Research (NWO) that is funded by the Dutch Ministry of Education, Culture and Science (OCW). We acknowledge financial support from an NWO-Graduate Program grant.

\newpage
\bibliographystyle{unsrt}
\bibliography{Bibliography}

\end{document}